\documentclass[aps,prd,preprint,groupedaddress,showpacs,eqsecnum,nofootinbib,floatfix]{revtex4}
\usepackage{graphicx,epsf,subfigure,amsfonts}

\graphicspath{{./plots/},{./diagrams/}}

 \newif\ifdraft
\drafttrue
\newif\ifpreprint
\preprinttrue

\def\fig#1{fig.~{\ref{#1}}}

\def\sect#1{section~{\ref{#1}}}
\def\eqn#1{eq.~(\ref{#1})}

\def\eqns#1#2{eqs.~(\ref{#1}) and~(\ref{#2})}
\def\eqns#1{eqs.~(\ref{#1})}

\def\ib{\bar\imath}
\def\jb{\bar\jmath}
\def\ms#1{\text{\small $#1$}}
\def\as#1{a_{\sigma(#1)}}
\def\q#1{ {q_{#1}}}
\def\qb#1{ {\overline{q}_{#1}}}
\def\Q#1{ {Q_{#1}}}
\def\Qb#1{ {\overline{Q}_{#1}}}

\newcommand{\be}{\begin{equation}}
\newcommand{\ee}{\end{equation}}
\newcommand{\bea}{\begin{eqnarray}}
\newcommand{\eea}{\end{eqnarray}}

\newcommand{\BlackHat}{{\sc BlackHat}}
\newcommand{\SHERPA}{{\sc SHERPA}}
\newcommand{\HELAC}{{\sc Helac-1Loop}}

\newcommand{\QGRAF}{{\sc QGRAF}}
\newcommand{\FORM}{{\sc FORM}}

\def\Tr{\mathop{\rm Tr}\nolimits}

\def\Wjjjj{$W\,\!+\,4$}

\def\Wjjjx{$W\,\!+\,1,2,3$}

\def\Wjn{$W\,\!+\,n$}

\begin{document}
\hfuzz 10 pt

\ifpreprint
\noindent
UCLA/11/TEP/110
\fi

\title{Colour Decompositions of Multi-quark One-loop QCD Amplitudes}

\author{Harald Ita$^{a,b,c}$ and Kemal Ozeren$^a$}

\affiliation{
\centerline{ $^a$Department of Physics and Astronomy, UCLA,}\\
\centerline{ Los Angeles, CA 90095-1547, USA}\\
\centerline{$^b$Niels Bohr International Academy and Discovery Center, NBI,}\\
\centerline{Blegdamsvej 17, DK-2100 Copenhagen, Denmark}
\centerline{$^c$Raymond and Beverly Sackler School of Physics and Astronomy,}\\
\centerline{Tel-Aviv University, Tel-Aviv 69978, Israel}
}

\date{\today}

\begin{abstract}
We describe the decomposition of one-loop QCD amplitudes in terms of
colour-ordered building blocks.  We give new expressions for the
coefficients of QCD colour structures in terms of ordered objects
called primitive amplitudes, for processes with up to seven
partons. These results are needed in computations of high-multiplicity
scattering cross sections in next-to-leading-order (NLO) QCD.  We
explain the origin of new relations between multi-quark primitive
amplitudes which can be used to optimise efficiency of NLO
computations.  As a first application we compute the full-colour
virtual contribution to the cross section for \Wjjjj{}-jet production
at the Large Hadron Collider, and verify that it is very well
approximated by keeping only the leading terms in an expansion around
the formal limit of a large number of colours.

\end{abstract}


\maketitle

\renewcommand{\thefootnote}{\arabic{footnote}}
\setcounter{footnote}{0}

\section{Introduction}
\label{IntroSection}

Scattering processes at the Large Hadron Collider (LHC) are dominated
by the strong interactions. Of particular interest in the search for
new physics are those which occur at high momentum transfer. These
processes frequently mimic new physics signatures, and so a solid
theoretical understanding is important to maximize the sensitivity of
experimental searches. Perturbative calculations in QCD, including
quantum corrections, provide first-principle predictions in this
direction, strongly linking fundamental theory to experiment.

Recent years have seen remarkable progress in perturbative QCD, with new
predictions at next-to-leading order (NLO) in the QCD coupling for many
processes of interest at the LHC~\cite{LH2010}. 
The present state-of-the-art studies of parton-level NLO QCD
processes include three~\cite{NLOJET,PRLW3BH,EMZW3Tev,W3jDistributions,TeVZ} or
four~\cite{BDDP,MStop,OtherUnitarityNLO,OPPNLO,W4jets,Z4jets,helactt2j}
coloured final states.  Such computations are challenging and rely on
efficient algorithms for organising the various dynamic degrees of freedom,
such as kinematics, spin and colour quantum numbers.  In this article we focus
on one particular aspect of these algorithms, namely the handling of the colour
degrees of freedom for loop computations. Our results are of immediate use in
modern approaches to NLO computations in QCD.

At tree level various approaches for dealing with colour have been developed
with differing advantages.  The colour-ordered approach groups together terms
with identical $SU(N_c)$ colour factors (where $N_c$ is the number of colours).
It was first developed for tree-level amplitudes in~\cite{BGColor,MColor} and
later extended for use in loop computations in
refs.~\cite{BKColor,Neq4Oneloop,TwoQuarkThreeGluon,Zqqgg}.  This approach
allows the symmetry properties
and simplified kinematic dependence of ordered amplitudes to be exploited.  
Alternatively, one can treat colour simultaneously with the kinematic
variables~\cite{BGcolordressed}. This colour-dressed method, most effectively
implemented as a Berends-Giele recursion~\cite{BGrelations}, has been shown to
be very efficient for tree-level computation and is naturally applied to
generic Standard Model processes~\cite{alpha,helactree,othertree,COMIX}.

At loop level the interaction of colour degrees of
freedom is more complex, and requires refined techniques.
For a large number of coloured final states the ordered approach has proved
particularly useful (see
e.g.~\cite{PRLW3BH,EMZW3Tev,W3jDistributions,W4jets,Z4jets}).  For this colour
organization an expansion in powers of $1/N_c$ is naturally set up which allows
for significant efficiency gains in the numerical phase space
integration~\cite{W3jDistributions}; for a fixed integration error, fewer
evaluations are needed for the smaller, but typically more time consuming,
subleading-colour contributions~\cite{DSColor}.

The colour-ordered approach is also naturally suited to modern methods
for loop computations using
unitarity~\cite{Neq4Oneloop,UnitarityMethod,DdimUnitarity} and on-shell
recursion~\cite{Bootstrap,CoeffRecursion,BlackHatI}. In particular,
generalised unitarity~\cite{Zqqgg,BCFUnitarity,Darren} and its
numerical extension~\cite{OPP,EGK,BlackHatI,GKM,GZRocket} was first developed
in the context of the colour-ordered approach. Ordered loop amplitudes have
simplified analytic properties and can be built through unitarity cuts from
ordered tree amplitudes. The division of the full amplitude into ordered pieces
also leads to improved numerical control.  By now a number of different methods
for dealing with colour are in use for computing hard virtual scattering matrix
elements~\cite{BlackHatI,OtherLargeN,GKW,SAMURAI,LoopAuto,helacnlo,BBU}.  This
includes colour sampling recursive approaches~\cite{GKW,OPPNLO}, in analogy to
those developed at tree level.  More traditional methods to organise loop
computations using colour information in the context of a Feynman-diagram
calculation have been applied in~\cite{FeynColor}.

Here we will follow the method based on colour-ordered amplitudes.  In
this approach the colour information is factorised from
gauge-invariant purely kinematic pieces, the latter being embodied in
primitive amplitudes~\cite{TwoQuarkThreeGluon}. Intermediate steps of
the computation are then free of any colour information. This
information is processed separately and reincorporated at the end of
the calculation.  As yet this approach is the only one which has led
to phenomenological applications with five final state
objects~\cite{W4jets,Z4jets}.  However, one technical issue with the
ordered approach is that the decomposition of matrix elements, and in
particular the virtual piece, in terms of primitive amplitudes, can be
non-trivial, especially as the number of fermion lines increases.  The
central difficulty is that the flow of quarks through a colour-ordered
diagram is not uniquely determined by the external colour charges;
each quark may navigate the loop in two distinct ways.  In addition,
the $1/N_c$ terms in the gluon propagator do not cancel when exchanged
between quark lines, as they do in the purely gluonic case.  In this
paper we show how the derivation of the colour decomposition can be
automated using Feynman diagrams. This only needs to be done once for
a given process, and is therefore not an issue for the numerical
efficiency of an NLO program.

At leading order in the number of colours, the decomposition in terms
of primitives is in general straightforward to write down.  The
corresponding results at subleading colour are non-trivial to derive,
in particular, when multiple quark lines are present; up to now they
have been dealt with on a case-by-case
basis~\cite{4qcolordec,Zqqgg,ZqqQQ,W3jcolordec}. Using our automated
approach we will derive explicit expressions at fixed multiplicity for QCD
amplitudes with many quark lines in terms of ordered objects. In particular, we
present new results for the decomposition of four- and six-quark amplitudes
with up to seven coloured states. (See ancillary text files for our explicit
expressions~\cite{partialsdata}.) Thus we provide the missing pieces for the
complete colour decomposition of QCD amplitudes with six and seven coloured
states. These new results are of much wider relevance, however.  With the
colour decomposition of the pure QCD amplitudes at hand one can obtain rather
simply their generalisations to include additional colourless objects such as
leptons, vector bosons and Higgs bosons.

As a state-of-the-art application of our results we present
distributions of the virtual contribution to \Wjjjj-jet production including
subleading-colour terms. A study at the LHC of this process has been published
in ref.~\cite{W4jets}. In that study the finite part of the virtual piece was
given at leading order in a large-$N_c$ colour expansion. Here we confirm that
the contribution of subleading-colour terms is small (a few percent), and shifts
the virtual contribution uniformly over phase space.

Finally, we discuss an additional result, that is certain relations
among the set of multi-quark primitive amplitudes.  These relations
appear as a by-product of our method and originate in the antisymmetry
of the colour stripped fermion-fermion-gluon vertex. While manifest at
low multiplicity, these ``fermion-flip'' relations become more and
more intricate as coloured states are added.  From a practical point
of view, such relations enable us to express one-loop amplitudes in
different ways. This can be exploited to obtain a minimal set of
contributing ordered amplitudes, allowing a reduction of the computing
resources needed for NLO calculations.

The organisation of this article is as follows. We begin
in~\sect{ColourDecomposition} by describing in detail our conventions
for dealing with the colour degrees of freedom. In~\sect{Setup} we
describe our algorithm and some details of our calculation. We also
discuss how relations between primitive amplitudes are obtained as a
by-product.  In~\sect{Results} we present selected results and a
numerical study of the virtual contribution to \Wjjjj{}-jet
production. Finally, we draw our conclusions and present an outlook for
future research.

\section{Colour Decomposition}
\label{ColourDecomposition}

In this section we review how full QCD one-loop amplitudes can be
decomposed into contributions associated with $SU(N_c)$ colour
structures.  While some of the notation extends the current
literature, most notation and concepts follow the detailed discussions
in~\cite{BKColor,TwoQuarkThreeGluon,Zqqgg} and the
reviews~\cite{TreeReview,OneLoopReview}. (See also ref.~\cite{EKMZreview}
for a recent review with topics related to this subject.)

\subsection{Basic conventions}
\label{sec:basic_colour}

We will be concerned with QCD scattering amplitudes of multiple quarks
and gluons.  Quarks are denoted by $q$ and $\bar q$ and are assumed to
transform in the fundamental $N_c$ and anti-fundamental $\bar N_c$
representations of the gauge group $SU(N_c)$.  When multiple quark
flavours appear\footnote{We will consider only amplitudes with
strictly different quark flavours. The cases with identical quarks are
easily obtained as linear combinations of these.}, they will be
distinguished by a flavour index $q_l$ and $\qb{l}$ with $l=1,2,\ldots
n_f$.  In the absence of this index, the quarks $q$ and $\qb{}$ are
understood to have a flavour number of one. We will keep the rank
$N_c$ as a free parameter and specialize to $N_c=3$ only for the
numerical results.  The colour indices of quarks and anti-quarks are
denoted by $\{i,j,\ldots\}$ and $\{\ib,\jb, \ldots\}$ respectively.
Gluons will be denoted by $g$ with adjoint colour indices labelled by
$a, b, c,$ etc.

It will be convenient to consider also fermions in the adjoint
representation of the gauge group. These will be denoted by capital
letters $Q$ and $\Qb{}$ and, in analogy to the fundamental fermions
above, by $Q_l$ and $\Qb{l}$ with an explicit flavour label. The
distinction of fermions and anti-fermions is then induced from the
flavour charge. As above the fermions $Q$ and $\Qb{}$ are understood
as flavour one states, serving as a shorthand notation for $\Q{1}$
and $\Qb{1}$.

\subsection{Colour algebra}

The aim of the colour decomposition is to disentangle colour and
kinematic degrees of freedom. This leads to a reorganization of
scattering amplitudes in terms of ordered subamplitudes, which are
typically easier to compute than the full amplitude itself. In
addition, colour decomposition gives control over the interplay of
colour and kinematic degrees of freedom. This can be used, for
example, for the expansion of scattering amplitudes in terms of powers
of the rank, $N_c$, of the gauge group $SU(N_c)$. Such an expansion
can be exploited for efficiency
gains~\cite{DSColor,W3jDistributions} in numerical NLO
computations.

In a Feynman-diagram representation of a scattering amplitude, gluon
interactions are proportional to the structure constants $f^{abc}$ of the
gauge group,
\be
[T^a, T^b]= i \sqrt{2} \,f^{abc}\,T^c,
\ee
where $T^a$ are the Lie algebra generators in the fundamental
representation, normalised such that $\Tr(T^a T^b)=\delta^{ab}$. The
quark-gluon interactions carry the fundamental representation
generators $(T^a)_i^{\,\jb}$.

We seek a uniform way of treating adjoint and fundamental colour
indices, allowing the identification of common group theory
factors. To this end we re-write all colour factors in terms of
fundamental representation generators $(T^a)_i^{\,\bar j}$ using
\be \label{eq:fabc} f^{abc} = -\frac{i}{\sqrt{2}} \Tr([T^a, T^b]\, T^c)\,.  \ee
Contracted adjoint indices can be reduced using the Fierz identity,
\be \label{eq:fierz} 
\sum_a\,(T^a)_{k}^{\,\,\ib} (T^a)_{l}^{\,\jb} =
\delta_{k}^{\,\,\jb}\,\delta_{l}^{\,\,\ib} - \frac{1}{N_c}
\delta_{k}^{\,\,\ib}\, \delta_{l}^{\,\,\jb}\,. 
\ee
All group theory factors can then be written in a canonical way in
terms of combinations of basic group theory data: powers of $N_c$,
traces $\Tr(T^{a_1} T^{a_2} \ldots T^{a_k})$, generator strings
$(T^{a_1} T^{a_2} \ldots T^{a_k})_i^{\,\jb}$ and Kronecker deltas
$\delta_i^{\jb}$. No repeated adjoint indices appear, as they are all
reduced using the Fierz identity~(\ref{eq:fierz}).

Within generic formulae, when labels take boundary values, it is useful to
adopt a slightly unconventional notation to keep expressions simple. A typical
case that appears is a colour structure,  $(T^{a_1} T^{a_2} \ldots
T^{a_k})_i^{\,\jb}$ with $k\rightarrow 0$. The natural interpretation for this
is to replace this colour structure by a Kronecker delta,
\be\label{def:boundary_colour_string}
(T^{a_1} T^{a_2} \ldots T^{a_k})_i^{\,\jb}\rightarrow \delta_i^{\jb}
\quad\mbox{for}\quad k=0\,.
\label{repl1}
\ee
Similarly, two boundary cases appear for traces of $SU(N_c)$ generators,
\bea\label{def:boundary_colour_trace}
\Tr(T^{a_1} T^{a_2} \ldots
T^{a_k})&\rightarrow&\Tr(T^{a_1})=0\quad\mbox{for}\quad k=1\,,
\label{repl2}\\ 
\Tr(T^{a_1} T^{a_2} \ldots T^{a_k})&\rightarrow&1\quad\mbox{for}\quad k=0\,.
\label{repl3}
\eea
One might find the replacement $\Tr(T^{a_1} T^{a_2} \ldots
T^{a_k})\rightarrow\Tr(1)=N_c$ more natural for $k=0$, however, we prefer to
absorb factors of $N_c$ into the definition of other objects, i.e. the partial
amplitudes as defined below in~\sect{sect:partialampl}.

The above replacement rules (\ref{repl1}), (\ref{repl2}) and (\ref{repl3}) will
be understood implicitly throughout the present article.

\subsection{Partial amplitudes}
\label{sect:partialampl}

Grouping terms of scattering amplitudes according to their colour structure
gives their decomposition into partial amplitudes.  The latter are in
one-to-one correspondence with the distinct colour structures that appear in a
scattering amplitude. 
In this section we spell out the form of two-quark, four-quark and six-quark
amplitudes, as they will be used in the later parts of the article. We consider
these amplitudes for distinct quark flavours. The cases with identical quark
flavours can be obtained by appropriate (anti-) symmetrization of quark
labels and we will not state them explicitly.

The partial amplitudes associated with the generic colour structures,
\be
\Tr(T^{a_1}\cdots T^{a_{j-1}})\,
(T^{a_{j+1}}\cdots T^{a_{j+k-1}})_{i_{j}}^{\,\,\ib_{j+k}}\,
(T^{a_{j+k+2}}\cdots T^{a_{j+k+l}})_{i_{j+k+1}}^{\,\,\ib_{j+k+l+1}}\cdots
\,,
\ee
will be denoted by,
\be
{A}_{n;j,k,l,...}(\ms{1,\ldots,j-1};
\ms{j_q,j+1,\ldots,(j+k)_{\qb{f_1}}};
\ms{(j+k+1)_{q_2},j+k+2,\ldots,(j+k+l+1)_{\qb{f_2}}};\ldots) \,,
\label{eq:partialampl}
\ee
mimicking the index structure of the colour traces and generator
strings.  As usual, legs without particle subscripts are taken to be
gluons.  The variables $f_i$ take values within the set
$\{1,\cdots,n_f\}$ according to the flavour arrangement of the
quarks. At tree level fewer colour structures appear; the single
colour trace is absent if at least one quark pair is present.  The
indices $j$ and $k,l,\ldots $ specify the size of colour traces and
generator strings, respectively.

Partial amplitudes have symmetry properties implied by the colour structures.
This includes a cyclic symmetry under rotation of gluons in a colour trace,
\be
A_{n;j,...}(1,2,\ldots,j-1;\ldots) = {A_{n;j,...}}(2,\ldots,j-1,1;\ldots)\,,
\ee
where we rotated the gluon indices associated with the colour
trace. The full symmetry group generated by this operation is
$\mathbb{Z}_{j-1}$.  Similarly, the order in which the colour
structures, $(T^{a_m}\cdots T^{a_n})_j\,^{\ib}$, appear leaves the value of
the partial amplitudes invariant,
\bea
&&{A}_{n;j,k,l,...}(\ldots; \ms{j_q,\ldots,(j+k)_{\qb{f_1}}};
\ms{(j+k+1)_{q_2},\ldots,(j+k+l+1)_{\qb{f_2}}};\ldots) =\\ &&\quad
{A}_{n;j,l,k,...}(\ldots; \ms{(j+k+1)_{q_2},\ldots,(j+k+l+1)_{\qb{f_2}}};
\ms{j_q,\ldots,(j+k)_{\qb{f_1}}}; \ldots) \,,\nonumber
\eea
and similarly for the exchange of other combinations of quark pairs.

Explicit decomposition of scattering amplitudes into partial amplitudes will be
given in the following.

\subsubsection{Two-quark partial amplitudes}

At tree level the two-quark QCD
amplitude~\cite{CLSColor,BGColor,MPXColor,BGngluons,MColor,MQKosower} is
\be \label{quarksubleading} 
\mathcal{A}_n^{{\rm tree}}(1_q,2_{\overline{q}},3,\dots,n) = 
\sum_{\sigma \in S_{n-2}}
(T^{a_{\sigma(3)}} \dots  T^{a_{\sigma(n)}})_{i_1}^{\,\,\ib_2} 
A_n^{{\rm tree}}(1_q,\ms{\sigma(3),\dots,\sigma(n)},2_{\overline{q}}),
\ee
where $S_{n-2}$ denotes the permutation group of $n-2$ gluon labels with the amplitudes
$A_n^{{\rm tree}}$  the tree level partial amplitudes. The elements of the
permutation group $\sigma$ are used in a two-fold way. On the one hand as a
permuted list of gluon labels and, on the other hand, as functions,
$\sigma(k)$, specifying the map of a given gluon, here gluon $k$, under the
permutation $\sigma$.  The colour structures here consist of a string of
fundamental generators, terminated with the indices of the quark and
anti-quark. We suppress the coupling constants here and throughout this
paper.

At one-loop level the colour decomposition into partial amplitudes was given for all
gluon multiplicities in ref.~\cite{TwoQuarkThreeGluon}.  In addition to the
tree-level colour structures, colour traces appear, 
\bea \label{eq:2q_colour_dec} 
\mathcal{A}_n^{{\rm}}(1_q,2_{\overline{q}},3,\dots,n) &=&\\ 
&&\hspace{-2.cm}\sum_{j=1}^{n-1}
\sum_{\ms{\sigma \in S_{n-2}/S_{(n-2;j)}}}\hspace{-.5cm}
\Tr(T^{\as{3}}\dots T^{\as{j+1}})\, (T^{\as{j+2}}\dots
T^{\as{n}})_{i_1}^{\,\,\ib_2}\nonumber\\
&&\hspace{1.0cm}\times
A_{n;j}(\ms{\sigma(3),\ldots,\sigma(j+1)};1_q,\ms{\sigma(j+2),\ldots,\sigma(n)},2_{\overline{q}}).
\nonumber 
\eea
The sum in \eqn{eq:2q_colour_dec} runs over all distinct colour
structures. This is achieved by a double sum. The outer sum runs over
the group theory structures and, within each of these, the inner sum
runs over independent gluon orderings.  Given the cyclic symmetry of
the colour traces $\Tr(T^{\as{3}}\dots T^{\as{j+1}})$, the full permutation
group $S_{n-2}$ of the $n-2$ gluons is reduced by the cyclic subgroups
$S_{(n-2;j)}\equiv\mathbb{Z}_{j-1}$ that leave the traces in
(\ref{eq:2q_colour_dec}) invariant. We adopt here the notation in
eqs.~(\ref{def:boundary_colour_string})
and~(\ref{def:boundary_colour_trace}) for the boundary cases $j=n-1$
and $j=1$, respectively.

\subsubsection{Four-quark partial amplitudes}

We can follow analogous steps for amplitudes with two pairs of
(distinct-flavour) quarks. Starting from a Feynman-diagram
representation of a $(q_1\, \qb{1}\, q_2\, \qb{2}\, g \dots g)$
amplitude, repeated application of eqs.~(\ref{eq:fabc}) and
(\ref{eq:fierz}) leads to the colour decomposition in terms of partial
amplitudes.

It is convenient to give the explicit form of the tree amplitudes for
four quarks and $n$ gluons~\cite{CLSColor,MColor,MQKosower,GGearlycolordec},
\bea
\mathcal{A}_n^{{\rm tree}}(1_{\q{1}}, 2_{\qb{1}},3_{\q{2}},4_{\qb{2}};5,\dots,n) &=&\\ 
&&\hspace{-5cm}
\sum_{k=1}^{n-3}
\sum_{\ms{\pi\in P_2}}
\sum_{\ms{\sigma \in S_{n-4}}}
(T^{a_{\sigma(5)}} \cdots T^{a_{\sigma(k+3)}})_{i_{\pi(1)}}^{\,\,\ib_2}
(T^{a_{\sigma(k+4)}} \cdots T^{a_{\sigma(n)}})_{i_{\pi(3)}}^{\,\,\ib_4}\nonumber\\
&&\hspace{-4cm}\times A^{\rm tree}_{n;k}(
\pi(1_\q{1}),\ms{\sigma(5),\ldots,\sigma(k+3)},2_\qb{1};
\pi(3_\q{2}),\ms{\sigma(k+4),\ldots,\sigma(n)},4_\qb{2}).\nonumber
\eea For the boundary cases $k=1$ and $k=n-3$ the colour structures
$(T^{a_{\sigma(5)}} \cdots
T^{a_{\sigma(k+3)}})_{i_{\pi(1)}}^{\,\,\ib_2}$ and
$(T^{a_{\sigma(k+4)}} \cdots
T^{a_{\sigma(n)}})_{i_{\pi(3)}}^{\,\,\ib_4}$ have to be replaced by
$\delta_{i_{\pi(1)}}^{\,\,\ib_2}$ and
$\delta_{i_{\pi(3)}}^{\,\,\ib_4}$ respectively (see
\sect{sec:basic_colour}).  $P_2$ stands for the permutation group of
the quarks $\{1_\q{1},3_\q{2}\}$.  In an overloaded notation, the
elements $\pi\in P_2$ are also interpreted as maps on the quarks and
quark labels, e.g. the notation $\pi(1_\q{1})$ and $\pi(1)$ is used to
denote mappings of quark and quark label, respectively.  The signs
keeping track of the sign in the $(-1/N_c)$-term (see
eq.~(\ref{eq:fierz})) in the colour-octet projector of the gluon
propagator are pulled into the definition of the partial amplitudes in
terms of colour-ordered tree amplitudes. Details can be found in the
supplied data files as described in~\sect{sect:datafiles}.

At one-loop level a similar colour decomposition into partial amplitudes gives~\cite{4qcolordec,ZqqQQ,W3jcolordec},
\bea \label{def:4q_colour_dec} 
\nonumber&&\mathcal{A}_n^{{\rm}}(1_\q{1},2_\qb{1},3_\q{2},4_\qb{2};5,\dots,n) =\\
&&\mathop{\sum_{\ms{j=1,n-3}}}_{\ms{k=1,n-(j+2)}} 
\sum_{\ms{\pi\in P_2}}\,
\sum_{\ms{\sigma \in S_{n-4}/S_{(n-4;j)}}}
\Tr(T^{a_{\sigma(5)}} \cdots T^{a_{\sigma(j+3)}})\nonumber\\
&&\hspace{1cm}\times
(T^{\as{j+4}} \cdots T^{\as{j+2+k}})_{i_{\pi(1)}}^{\,\,\ib_2}
(T^{\as{j+3+k}} \cdots T^{\as{n}})_{i_{\pi(3)}}^{\,\,\ib_4}\\
&&\hspace{1cm}\times
A_{n;j,k}(\sigma_{(5,j+3)};
\pi(1_{q_1}),\sigma_{(j+4,j+2+k)},2_{\qb{1}};
\pi(3_{q_2}),\sigma_{(j+3+k,n)},4_{\qb{2}})
\,,\nonumber
\eea
where we again use the conventions for the boundary cases given in
\sect{sec:basic_colour}. We also introduced the abbreviation
$\sigma_{(i,j)}=\ms{\{\sigma(i),\sigma(i+1),\ldots,\sigma(j-1),\sigma(j)\}}$.

\subsubsection{Six-quark partial amplitudes}

We now consider amplitudes with three distinct quark flavours.
Starting from a Feynman-diagram representation of a $(q_1\, \qb{1}\,
q_2\, \qb{2}\,q_3\,\qb{3}\,g \dots g)$ amplitude, repeated application
of \eqns{eq:fabc} and (\ref{eq:fierz}) once again leads to the colour
decomposition in terms of partial amplitudes. For simplicity of
exposition we give the six-quark and seven-quark one-gluon partial
amplitudes.

The explicit form of the tree amplitudes for six quarks is
given by~\cite{CLSColor,MColor,MQKosower},
\bea\label{eq:6qtree}
&&\mathcal{A}_6^{{\rm tree}}(
1_{\q{1}},2_{\qb{1}},
3_{\q{2}},4_{\qb{2}},
5_{\q{3}},6_{\qb{3}}
) = 
\sum_{\ms{\pi\in P_3}}
\delta_{i_{\pi(1)}}^{\,\,\ib_2}
\delta_{i_{\pi(3)}}^{\,\,\ib_4}
\delta_{i_{\pi(5)}}^{\,\,\ib_6}\\
&&
\hspace{3cm}
\times\, A^{\rm tree}_{6;1,1}(
\pi(1_\q{1}),2_\qb{1};
\pi(3_\q{2}),4_\qb{2};
\pi(5_\q{3}),6_\qb{3})\,,
\nonumber
\eea with $P_3$ the permutation group of the quarks
$\{1_\q{1},3_\q{2},5_\q{3}\}$.  

For one-loop amplitudes a similar colour decomposition into partial amplitudes gives,
\bea \label{def:6q_colour_dec} 
&&\mathcal{A}_6(
1_{\q{1}},2_{\qb{1}},
3_{\q{2}},4_{\qb{2}},
5_{\q{3}},6_{\qb{3}}
) = 
\sum_{\ms{\pi\in P_3}}\,
\delta_{i_{\pi(1)}}^{\,\,\ib_2}
\delta_{i_{\pi(3)}}^{\,\,\ib_4}
\delta_{i_{\pi(5)}}^{\,\,\ib_6}\\
&&\hspace{3cm}\times\, A_{6;1,1,1}(
\pi(1_\q{1}),2_\qb{1};
\pi(3_\q{2}),4_\qb{2};
\pi(5_\q{3}),6_\qb{3})\,.
\nonumber
\eea

The generalization of the above colour decompositions (\ref{def:6q_colour_dec})
to seven-point partial amplitudes is given by,
\bea \label{def:1g6q_colour_dec} 
&&\mathcal{A}_7( 1_{\q{1}},2_{\qb{1}}, 3_{\q{2}},4_{\qb{2}},
5_{\q{3}},6_{\qb{3}};7) = 
\sum_{\ms{\sigma\in \mathbb{Z}_3}}\, \sum_{\ms{\pi\in P_3}}\,
(T^{a_7})_{i_{\pi(1)}}^{\,\,\ib_{\sigma(2)}}\,
\delta_{i_{\pi(3)}}^{\,\,\ib_{\sigma(4)}}
\delta_{i_{\pi(5)}}^{\,\,\ib_{\sigma(6)}}
\\ && \hspace{3cm}
\times\,A_{7;1,2,1}( \pi(1_\q{1}),7,\sigma(2_\qb{1}); \pi(3_\q{2}),\sigma(4_\qb{2});
\pi(5_\q{3}),\sigma(6_\qb{3}))\,.  \nonumber 
\eea 
The cyclic group $\mathbb{Z}_3$ is understood to rotate the three quarks
$\{1_{\q{1}},3_{\q{3}},5_{\q{5}}\}$ in order to guarantee that the gluon is emitted from all quark sources.  Refined definitions of
various elements in \eqn{def:1g6q_colour_dec} follow conventions in~\eqn{eq:6qtree}.

Tree amplitudes take an analogous form, as
was the case for the 6-point formulas (\ref{def:6q_colour_dec}) and
(\ref{eq:6qtree}), so we do not state the tree-level formula here explicitly.

\subsection{Primitive amplitudes} \label{sect:PrimitiveAmpls}

Partial amplitudes can be further expressed as linear combinations of
`primitive amplitudes'~\cite{TwoQuarkThreeGluon}. In fact, primitive
amplitudes are defined as a set of gauge-invariant, colour-ordered
building blocks which suffice to express a given set of partial
amplitudes. We discuss in this subsection how to characterize and
generate primitive amplitudes.  Our algorithm for expressing partial
amplitudes in terms of primitive amplitudes will be given in
\sect{Setup}.

\begin{figure} \includegraphics[scale=0.5]{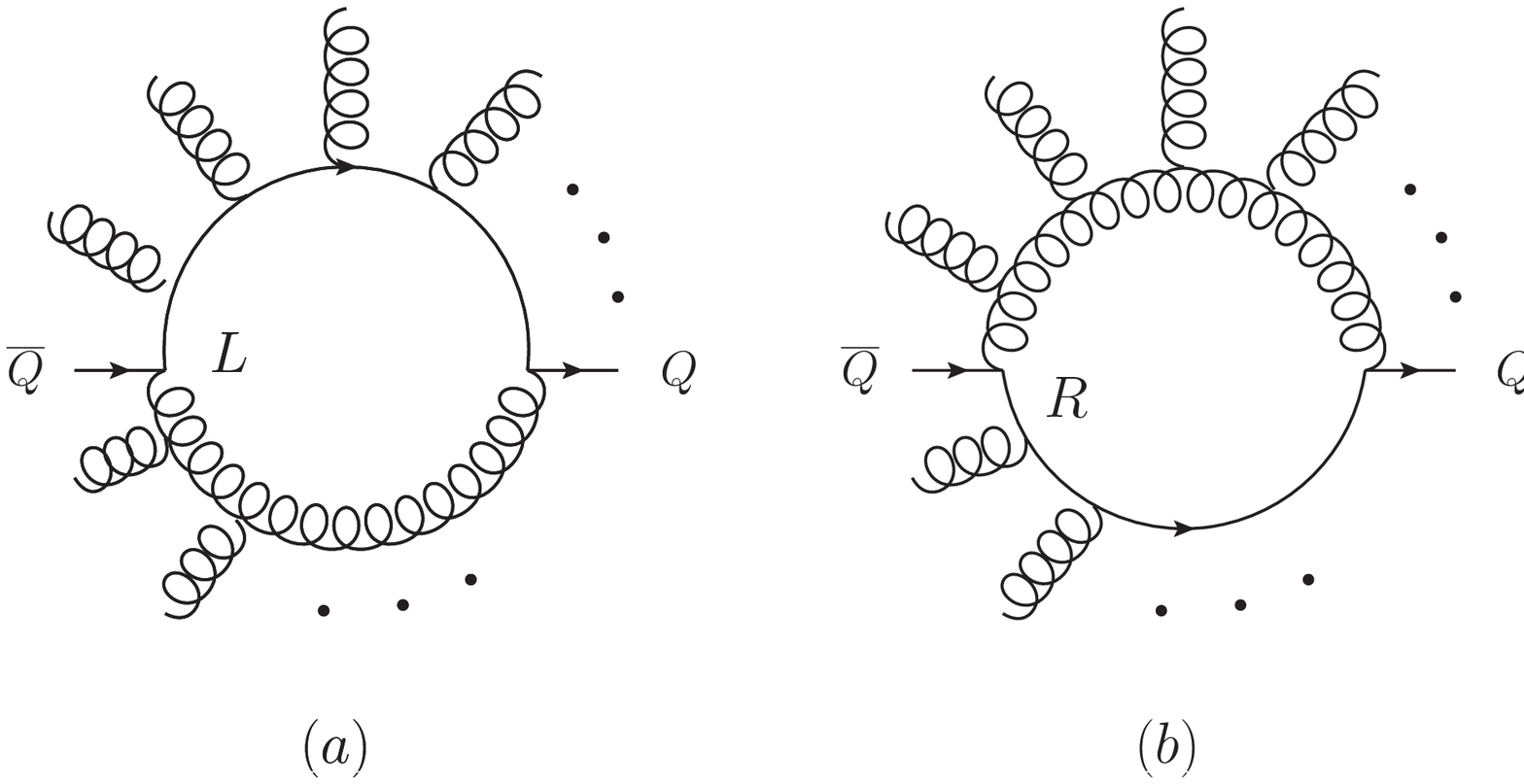}\\\vskip 0.3 cm
	\includegraphics[scale=0.5]{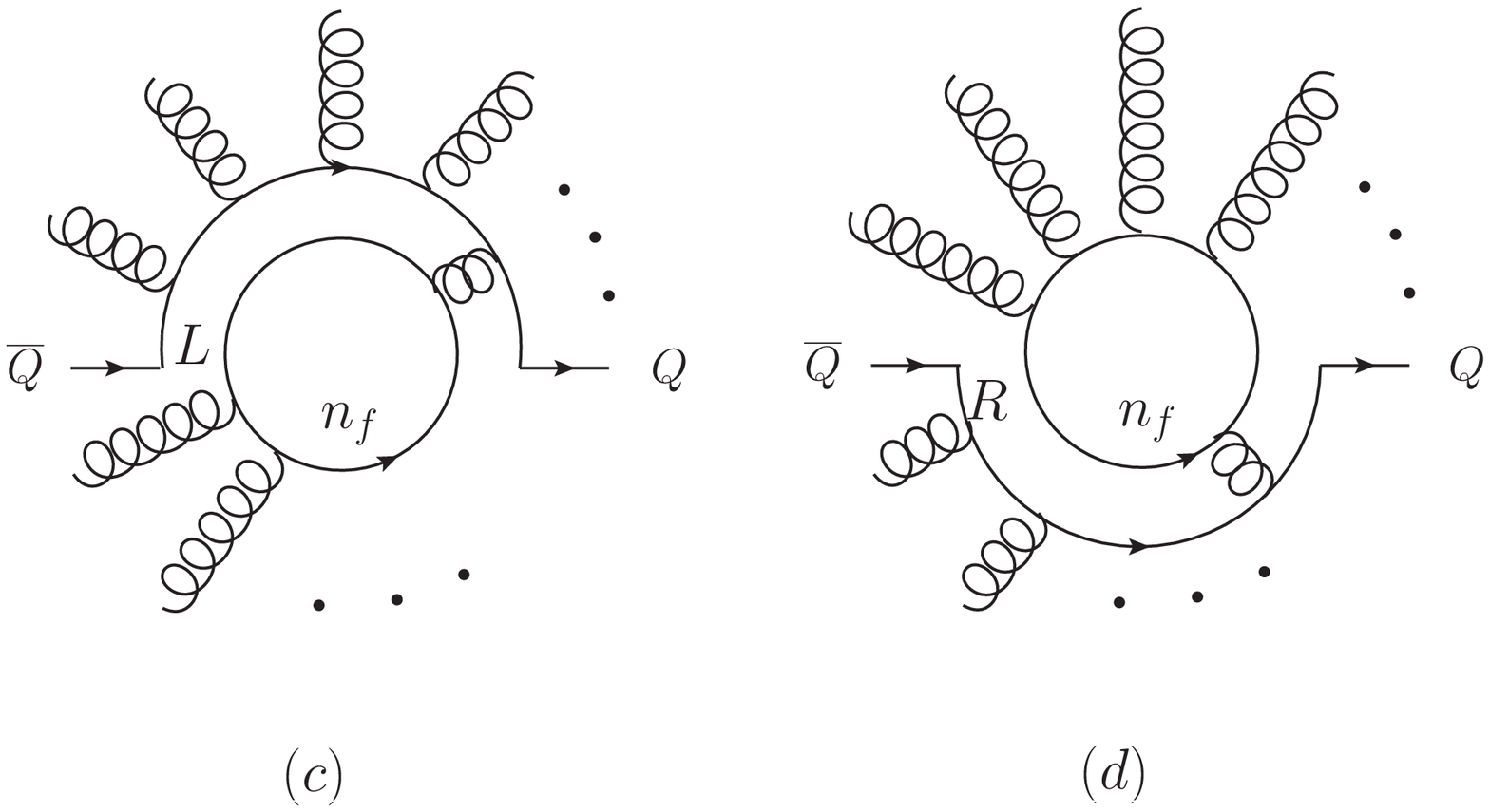}\\\vskip 0.3 cm
	\includegraphics[scale=0.5]{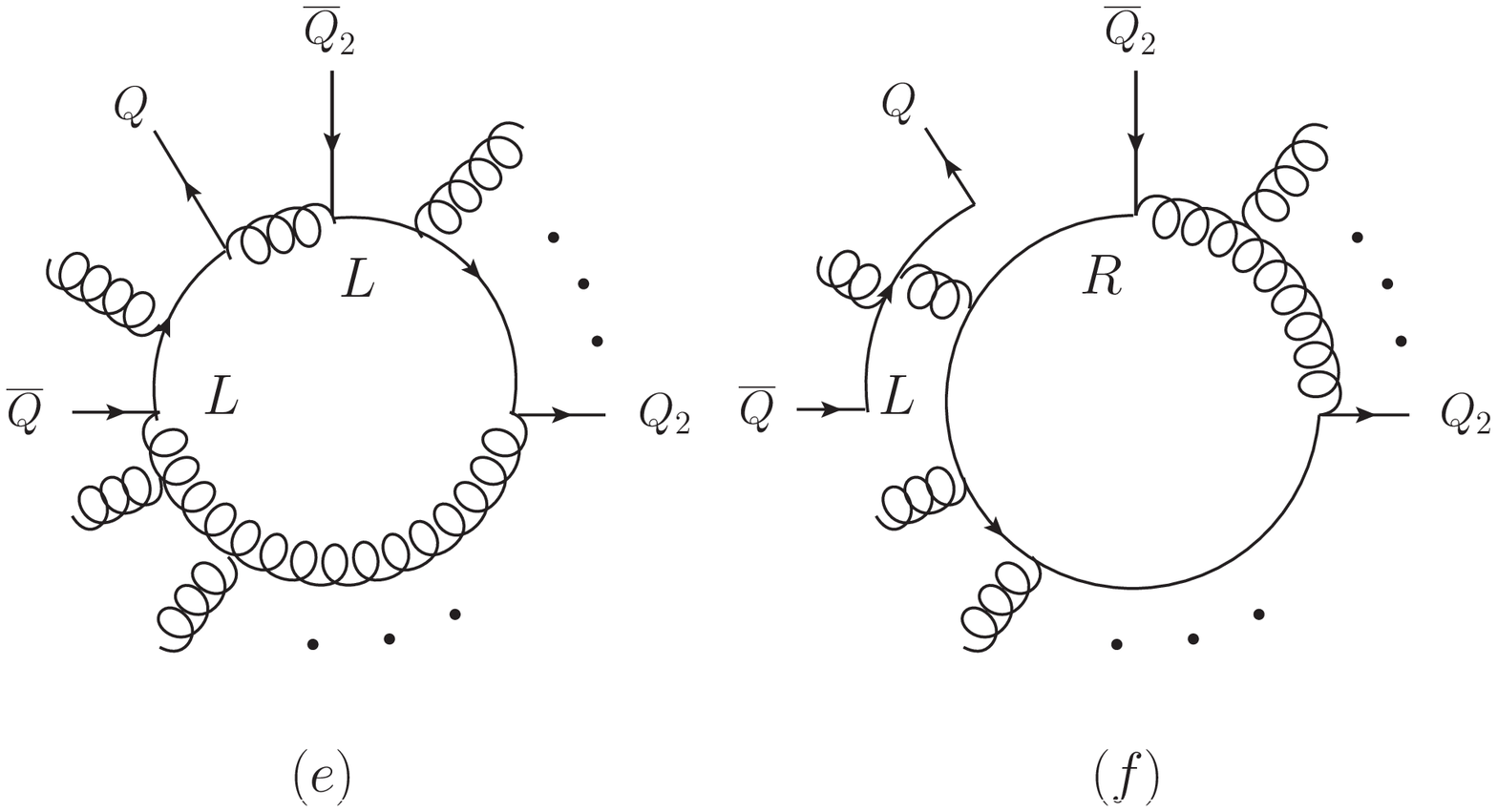}
\caption{ Parent diagrams with a distinct routing of quark lines: (a)
a parent with a `left-turner' fermion labelled by `$L$', (b) a
`right-turner' fermion, labelled by `$R$', (c) `$n_f$-left-turner'
($n_fL$) fermions, (d) `$n_f$-right-turner' ($n_fR$) fermions. Two
routing labels are associated to the distinct quark flavours $Q$ and
$\Q{2}$: (e) `left-left-turner' ($LL$) fermions, (f)
`left-right-turner' ($LR$) fermions. Note that in (f) the first
fermion does not in fact enter the loop. In general as many
$L/R$-labels appear as there are distinct external quark flavours.
Quarks are associated with capital letter $Q$'s, as in primitive
amplitudes they arise from adjoint representation colour assignments
in our algorithm.}
\label{fig:LTnfprimitive} \end{figure}

Primitive amplitudes can be specified by sets of adjoint representation
colour-ordered Feynman diagrams~\cite{TwoQuarkThreeGluon,OneLoopReview}. These
diagrams differ from the standard Feynman diagrams in that colour labels are
dropped. At gluon vertices, for example, the structure constants $f^{abc}$ are
omitted.  Gluon vertices dressed with colour labels are symmetric under the
exchange of legs.  Due to the anti-symmetry of the structure constants
$f^{abc}$, the remaining kinematic part is necessarily anti-symmetric under
exchange of legs.  In colour-ordered Feynman diagrams one must preserve the
ordering of the external legs to keep track of the signs associated with the
vertices. Given a set of colour-ordered Feynman diagrams one uses
colour-ordered Feynman rules (see e.g.~\cite{TreeReview}) to obtain amplitudes
and eventually primitive amplitudes.

We require the set of all primitive amplitudes needed to compute a given
scattering amplitude. In particular, we will need their representation in terms
of sets of Feynman diagrams.  This may be generated systematically in two
steps, which we will describe in full below: 
\begin{enumerate} 
\item{generate `coarse' primitive amplitudes,}
\item{split the coarse primitives according to the routing of fermions
around the loop. }
\end{enumerate}
The coarse primitives can be thought of as sums of planar, ordered diagrams,
obtained using colour-ordered Feynman rules. Each coarse primitive amplitude
corresponds to a particular ordering of the external legs.

In practise, we obtain the coarse primitive amplitudes starting from a
colour-dressed representation of the amplitude by taking all partons,
including quarks, in the adjoint representation of the gauge
group. Performing all the colour algebra reductions, one finds a set
of terms, some with single and some with double traces of fundamental
generators, just as is the case for purely gluonic amplitudes. The
coarse primitive amplitudes
\be
A^{\rm coarse}(1,\ldots,\Q{1},\ldots,\Qb{1}\cdots)\,,
\ee
are then identified as the coefficients of the single trace structures
\be N_c\,\Tr(T^{a_1} \cdots T^{a_{\Q{1}}}\cdots
T^{a_{\Qb{1}}}\cdots )\,.
\ee

Secondly, a finer splitup of the coarse primitive amplitudes into
actual primitive amplitudes is obtained by analysing the fermions'
routing around the loop. This can be considered a generalisation of
the familiar case of diagrams with a fermion circulating in the loop
not mixing under gauge transformations with diagrams with a gluon in
the loop.  Similarly, fermion routing information in Feynman diagrams
allows to split coarse primitive amplitudes into finer gauge-invariant
subsets.  The routing information of a fermion line specifies whether
the fermion turns left or right upon entering the loop. In the case
where the fermion does not enter the loop, the left/right label is
assigned according to which side of the loop the fermion passes. We
follow the fermion line starting from the anti-quark. In this way,
each fermion line is labelled as a left- or right-turner, denoted by
capital labels $L$ and $R$, respectively.  Some examples can be found
in~\fig{fig:LTnfprimitive}.

Algorithmically, the routing information of a colour-ordered Feynman
diagram can be determined by cutting a single loop propagator. If no
gluon propagator is available, the Feynman diagram has to correspond
to an $n_f$ term. We then cut the internal fermion line. If a gluon
loop-propagator is available we instead cut that one.  After cutting
the propagator the Feynman diagram corresponds to a diagram of a
colour-ordered tree amplitude. The fermion routing information is then
in one-to-one correspondence to the ordering of external fermions
relative to their anti-fermions and may be read off directly.  The
coarse primitive amplitudes can then be sorted diagram by diagram into
the finer classes of primitive amplitudes with definite fermion flow.

The explanation of the gauge invariance of the primitive amplitudes as
defined above has been given in ref.~\cite{TwoQuarkThreeGluon}.  The
basic reasoning relies on a non-standard colour charge assignment to
quarks and gluons.  Primitive amplitudes of standard QCD can typically
be interpreted as partial amplitudes of a theory with non-standard
gauge groups and colour charges. Gauge invariance of a class of
Feynman amplitudes is then ensured from the associated gauge
invariance of partial amplitudes.  The non-standard colour assignments
include various bifundamental and adjoint representations in product
gauge groups $SU(N_1)\times SU(N_2)\times SU(N_3)\ldots $ for quarks
and gluons respectively. The basic logic is only a minor
generalisation of that presented in the original
literature~\cite{TwoQuarkThreeGluon}, to which we refer for further
details.

\subsubsection{Notation} \label{sect:notation}

Pictorially, primitive amplitudes can be characterized by a small set of
`parent' Feynman diagrams. These are the representative colour-ordered Feynman
diagrams of a primitive amplitude with the maximal number of loop propagators.
An additional requirement is that the complete set of Feynman diagrams
associated with a primitive amplitude can be obtained from the parent diagrams
through pinching of propagators and `pulling out' of tree amplitudes.

Example parent diagrams of six distinct primitive amplitudes are shown in
\fig{fig:LTnfprimitive}. The diagrams are representative of classes of
colour-ordered Feynman diagrams.  Parent diagrams (a), (b), (c) and (d) have
identical external states, as do diagrams (e) and (f). They differ by the
routing of the fermion lines around the loop and are gauge invariant
individually\footnote{More precisely, the primitive amplitudes they correspond
to are gauge-invariant.}. The routing information of the fermions is given in
terms of left- and right-turner labels; $L$ and $R$. Closed fermion loops are
specified by an additional label $n_f$. Such routing labels must be specified
for each fermion line.

So instead of specifying a primitive amplitude by the set of colour-ordered
Feynman diagrams contributing to it, we can be more concise, and specify only
1) an ordered set of external states and 2) routing data of fermion lines.  

We will use this information to specify primitive amplitudes. In our
conventions, for the amplitudes shown in~\fig{fig:LTnfprimitive} we will use
the following notation, 
\bea
\nonumber(a):&&A^L(Q, \ldots g, g, \Qb{},g,g,g,g, \ldots )\,,\\
\nonumber(b):&&A^R(Q, \ldots g,g, \Qb{},g,g,g,g, \ldots )\,,\\
\nonumber(c):&&A^{n_fL}(Q, \ldots g, g, \Qb{},g,g,g,g,\ldots )\,,\\
\nonumber(d):&&A^{n_fR}(Q, \ldots g, g, \Qb{},g,g,g,g, \ldots )\,,\\
\nonumber(e):&&A^{LL}(Q, \Qb{2},g,\Q{2},\ldots g, g, \Qb{},g, \ldots )\,,\\
\nonumber(f):&&A^{LR}(Q, \Qb{2},g,\Q{2},\ldots g, g, \Qb{},g, \ldots )\,.
\eea
Fermion lines will be given ascending flavour numbers starting from $1$ (we
identify $Q \equiv Q_1$).  The routing information will be
specified by a list of labels $L$, $R$. $L$ stands for a fermion that turns
left upon approaching the loop, while $R$ stands for a right turning fermion.  The
first label in the list corresponds to the first flavour-line, the second label
to the second line etc. Closed fermion loops are indicated by an additional
label $n_f$. External states are given in clockwise order and we exploit the
rotational symmetry (described in the next section) to rotate quarks with
flavour one into the first position.

\subsubsection{Symmetry properties}

Pure-QCD primitive amplitudes enjoy a variety of symmetries.  These are useful
for reducing the number of independent primitive amplitudes that need to be
computed. We use the symmetries below to write primitives in a standard form
starting with a left-turner label, with the quark of flavour one in the first
position of the particle labels, 
\be
A^{L\ldots}(Q,\ldots )\quad\mbox{or}\quad A^{n_fL\ldots}(Q,\ldots )\,.
\ee 
The relevant symmetry transformations are described below.  In addition, an
extended set of relations between primitive amplitudes will be discussed below
in \ref{relations}.

First of all, cyclic rotation of the labels does not change the primitive
amplitudes,
\be A^{\ldots}(1,2,\ldots,n)=A^{\ldots}(2,\ldots,n,1).  \label{eqn:cyclsymm}
\ee

Another symmetry arises from flipping over the colour-ordered diagrams of a
given primitive. This reverses the ordering of the external states and
exchanges all left-and right-turners up to a sign. In the case of a two
quark line amplitude we have, 
\bea \label{eqn:flipsymm}
&&A^{LR}(1_q,2,\ldots,k_{\qb{}},\ldots,l_{\q{2}},\ldots,m_{\qb{2}},\ldots,n)=\\
&&\quad(-1)^n\,
A^{RL}(1_q,n,\ldots,m_{\qb{2}},\ldots,l_{\q{2}},\ldots,k_{\qb{}},\ldots,2)\,.\nonumber
\eea

A further generic relation between primitive amplitudes originates in the
reversal of a fermion's arrow. This leads to a flip of the routing label,
\be A^{\ldots L_i\ldots}(\ldots ,k_\q{i},\ldots,l_{\qb{i}},\ldots,n)= A^{\ldots
R_i\ldots}(\ldots ,k_{\qb{i}},\ldots,l_{\q{i}},\ldots,n).
\label{eqn:flipfermarrow} \ee
This property is naturally generalised to the case with multiple quark lines.
Each quark line may be reversed and the associated turner label flipped,
$L \leftrightarrow R$. We will not use this transformation here,
preferring to keep the distinction between quarks and anti-quarks.

\subsection{Colour decomposition of partial amplitudes.}
\label{sect:PartialNotation}
\begin{figure}[b]
	\includegraphics[scale=1.7,angle=-90]{diagrams/4q_1.dat.epsi}
	\caption{Four quark partial amplitude decomposed into primitive
	amplitudes, in the notation of the attached file.} \label{fig:FourQuarkPartial} \end{figure}

In this section we present our notation for partial amplitudes given as linear
combinations of primitives. As the expressions are lengthy we prefer to write
them in a simplified form, which is also more convenient for use with a
computer program. We display only a small sample in the main body of this
paper, and attach the remainder as text files~\cite{partialsdata}. Additional
information concerning the supplied data-files can be found below
in~\sect{sect:datafiles}.

To explain our notation we refer to \fig{fig:FourQuarkPartial} where we show
our result for the four quark amplitude.  The line starting with `* Partial'
indicates the beginning of the expression and is followed by a specification of
which particular partial amplitude is being considered. The notation for this
mimics the argument of the partial amplitudes (\ref{eq:partialampl}). It takes
the form of a comma separated list with each element of the list denoting an
$SU(N_c)$ colour structure. Such a colour structure is unambiguously specified
by a list of particles. Quarks in the fundamental representation are denoted by
$q$, and may have an additional numerical label to distinguish different
flavours ($q$, $q2$, $q3$ \dots). Anti-quarks are written $qb$, $qb2$, $qb3$
etc.  Momentum labels appear in parentheses, e.g. $q(1)$ denotes a quark that
carries the momentum $k_1$.  For the case of \fig{fig:FourQuarkPartial}, the
colour structure is $\delta_{i_1}^{\ib_2} \delta_{i_3}^{\ib_4}$.

There follows specification of the born contribution to this partial amplitude
in terms of colour-ordered tree-level amplitudes. The line starting with the
label,
\bea
\nonumber{\rm born}   &&\{\ldots |\ldots \\
\nonumber &&\ldots |\ldots \\
\nonumber &&etc. \\
\nonumber &&\} \,
\eea
shows a list of particles in the adjoint representation and, separated by a
vertical line, `$|$', the coefficient of the given tree within the partial
amplitude. Labels with capital letters $Q$, $Q2$ etc. are used for these kind
of quarks.  For partial amplitudes that are a linear combination of
colour-ordered tree amplitudes, multiple line entries are used, each displaying
a list of particle labels and the weight of the associated tree amplitude
within the partial amplitude.

In more conventional form the four quark born amplitude is given by,
\be
\mathcal{A}_4^{{\rm tree}}(1_q,2_{\qb{}},3_{\q{2}},4_{\qb{2}}) = A^{\rm
tree}(1_Q,2_{\Qb{}},3_{\Q{2}},4_{\Qb{2}}) \times\left(\delta_{i_1}^{\ib_4}
\delta_{i_3}^{\ib_2} -\frac{1}{N_c}\, \delta_{i_1}^{\ib_2}
\delta_{i_3}^{\ib_4}\right)\,.  \label{eqn:TreeAmpl}
\ee
The displayed partial amplitude corresponds to the second term
in~\eqn{eqn:TreeAmpl} and the ordered born amplitude is $A^{\rm
tree}(1_Q,2_{\Qb{}},3_{\Q{2}},4_{\Qb{2}})$ weighted by a factor of
$-\frac{1}{N_c}$.

The remaining entries specify the one-loop partial amplitude in terms of
primitives. This entry starts with the header `$loop$', 
\bea {\rm loop}
\nonumber&&\{\\
\nonumber&&\ldots| \ldots |\ldots \\
\nonumber&&\ldots| \ldots |\ldots \\
\nonumber&&{\rm etc.}\\
\nonumber\}\quad && \,
\eea
with the individual lines each specifying the turner labels,
an ordered list of external states and the weight of the primitive amplitude.
In total this represents a sum of primitive amplitudes, each with a $N_c$-
and/or $n_f$-dependent coefficient.  For the present example, displayed in
\fig{fig:FourQuarkPartial}, the partial amplitude is given by,
\bea
\mathcal{A}_4(1_q,2_{\qb{}},3_{\q{2}},4_{\qb{2}}) &=& \ldots +\,
A_{4;1,1}(1_q,2_{\qb{}};3_{\q{2}},4_{\qb{2}})\, \delta_{i_1}^{\ib_2}
\delta_{i_3}^{\ib_4}+\ldots\,,\\
A_{4;1,1}(1_q,2_{\qb{}};3_{\q{2}},4_{\qb{2}})&=&
\frac{1}{N_c^2}\,A^{LL}(1_Q,4_{\Qb{2}},3_{\Q{2}},2_{\Qb{}})+
\left(1+\frac{1}{N_c^2}\right)\,A^{LR}(1_Q,3_{\Q{2}},4_{\Qb{2}},2_{\Qb{}})\nonumber\\
&&\null +\frac{1}{N_c^2}\,A^{LR}(1_Q,4_{\Qb{2}},3_{\Q{2}},2_{\Qb{}})-
\frac{1}{N_c^2}\,A^{LL}(1_Q,2_{\Qb{}},4_{\Qb{2}},3_{\Q{2}})\nonumber\\ &&\null
-\frac{n_f}{N_c}\,A^{n_fLL}(1_Q,4_{\Qb{2}},3_{\Q{2}},2_{\Qb{}})\,.
\label{eqn:LoopAmpl}
\eea
The five primitive amplitudes appear in the same order as in
\fig{fig:FourQuarkPartial}, where they are given on separate
lines. Each primitive amplitude is specified by a list of particles in
the adjoint representation and string of fermion left/right-turner
labels. We denote adjoint quarks with a capital $Q$. Each quark line
has a `turner label' $L$ or $R$, which described the direction it
turns upon approaching the loop, as described in
\sect{sect:PrimitiveAmpls}.

The above expression can be compared to the one given for the partial amplitude
denoted by $A_{6;2}$ in
eq.(2.15) of ref.~\cite{Zqqgg}. To simplify this comparison we set the
number of scalars $n_s=0$ and drop the top-quark contributions in the
expression for $A_{6;2}$. We can omit the colour-neutral lepton
states, as they play no role in this colour decomposition. To match
the expression for $A_{6;2}$ with our \eqn{eqn:LoopAmpl} we have to
replace $2\leftrightarrow4$ in the expression for $A_{6;2}$. We can
then identify $A^{+-}(1,3,4,2)= A^{LR}(1,3,4,2)$,
$A^{++}(1,4,3,2)=A^{LL}(1,4,3,2)$, $A^{sl}(4,3,1,2)=A^{LL}(1,2,4,3)-
A^{LR}(1,4,3,2)$ and $A^{s,++}(1,4,3,2)+
A^{f,++}(1,4,3,2)=-A^{n_fLL}(1,4,3,2)$. (Compared to~\cite{Zqqgg} we
differ here by an overall sign in the definition of the $n_f$-terms,
which introduces this apparent relative minus-sign.)  This
identification can also be inferred from comparison of parent
diagrams. The relative minus sign in the expression for
$A^{sl}(4,3,1,2)$ appears from the asymmetry of the
fermion-fermion-gluon three-point vertex. Upon reversing the order of
the quarks $1_q$ and $2_{\qb{}}$ and rotation of the labels one finds
the identity, $-A^{LL}(1,2,4,3)=A^{RL}(2,1,4,3)=A^{RL}(1,4,3,2)$, with
the relative minus sign absorbed into the re-ordering of external
legs.

With this comparison we conclude the explanation of our conventions.

\section{Colour decomposition algorithm}
\label{Setup}

In this section we will describe our setup for obtaining partial
amplitudes in terms of primitive amplitudes, for arbitrary
processes. The algorithm we use is based on analysing Feynman diagrams
using their colour information. We identify partial amplitudes 
in terms of linear combinations of Feynman
diagrams. Similarly, we express primitive amplitudes in terms of
Feynman diagrams. Finally, we express partial amplitudes in terms of
primitive amplitudes by solving the linear set of equations to
eliminate the explicit dependence on the diagrams. In addition to the
colour decomposition of the partial amplitudes, we find non-trivial
relations between primitive amplitudes, which arise due to the
redundancy of the linear equations we solve.

\subsection{Setup}
We begin by generating all one-loop Feynman diagrams using
\QGRAF{}~\cite{QGRAF}. We remove diagrams with contact terms
(four-point gluon vertices), tadpoles and those with bubbles on
external lines. Let us suppose there are $N_d$ diagrams after this
procedure. The output of this program is then processed using the
computer algebra package \FORM{}~\cite{FORM}. Using this package we
dress the diagrams with colour - gluons in the adjoint and quarks in
the fundamental representation of the gauge group - and simplify the
colour algebra using eqs.~(\ref{eq:fabc}) and (\ref{eq:fierz}). There
is no need to substitute the kinematic parts of the Feynman rules, so
we leave this information implicit.  This procedure gives us an
expression for the amplitude in terms of sums of diagrams multiplied
by the various colour structures that may appear,
\be \mathcal{A}(\{ {\bf a}_g,{ \bf i}_q\}) = \sum_{i=1}^{N_P} A_i^{\rm partial}
C_i(\{ {\bf a}_g, {\bf i}_q \}).  \ee
On the left hand side is the fully colour-dressed amplitude, which
depends on the set of colour charges $\{\bf{a}_g,\bf{i}_q \}$ of the
external particles. The right hand side is its decomposition in terms
of $N_P$ partial amplitudes $A_i^{\rm partial}$, multiplying colour
structures $C_i$. Partial amplitudes and their colour structures are
discussed further in~\sect{sect:partialampl}.  Here and in the
following we leave implicit all dependence on momenta and helicities.

After processing the diagrams in this way, we find expressions
for the $A_i^{\rm partial}$ as sums of diagrams,
\be \label{partialamp} A_i^{\rm partial} = \sum_{j=1}^{N_d}  K_i\,^j\, d_{j}\quad\mbox{for}\quad i=1,\ldots,N_P.  \ee
Here $K_i\,^j$ is a $N_P \times N_d$ matrix which describes how the
$j$'th diagram contributes to the $i$'th partial amplitude. The
entries of the matrix $K_i\,^j$ consist of simple functions of the
rank of the gauge group, $N_c$.  The $d_i$ are the kinematic parts of
the set of Feynman diagrams, but we will not need their explicit form
here. It is sufficient to think of them as tags for the individual
Feynman diagrams.

With this in hand, we now discuss how the primitive amplitudes are
expressed in terms of diagrams. We first go back and dress the
original one-loop Feynman diagrams with colour assuming that all
external particles, including quarks, live in the adjoint
representation of the gauge group. Upon simplifying the colour
algebra, we find two types of term --- those with single traces, and
those with double traces, exactly as one finds for purely gluonic
amplitudes,
\be \mathcal{A}^{\rm adjoint}(\{{\bf a}_g,{\bf a}_q\}) = \sum_{\sigma \in
S_n/\mathbb{Z}_n} N_c \Tr(T^{a_{\sigma(1)}} T^{a_{\sigma(2)}} \cdots
T^{a_{\sigma(n)}})\, A^{\rm coarse}(\sigma) + \textrm{double trace terms.} \ee
In this equation we have added the label `adjoint' to remind the
reader that this is the amplitude with all external particles in the
adjoint representation. The sum is over the permutation group $S_n$ of
$n$ elements modulo cyclic rotations $\mathbb{Z}_n$. Some permutations
may not appear when fermion lines of the diagrams cannot be arranged
in a planar way.  We drop all double-trace terms. Note that there is
no approximation being made here --- this is simply the definition of
the primitive amplitudes.  (The double trace contributions can be
fully reconstructed from the primitive amplitudes, though we will not
need to do this here.)

If we were to restrict ourselves to the case of amplitudes with only adjoint
particles, then we could now identify the coefficients $A^{\rm
coarse}(1,2,\dots,n)$ of the single trace structures as primitive amplitudes, 
\be \label{coarseprimitiveamp} A_k^{\rm coarse}(\sigma_k) = \sum_{j=1}^{N_d}
(L^{\rm coarse})_k\,^j\, d_j, \qquad k=1,\ldots ,N_{\rm prim}^{\rm coarse}\,, \ee
where $(L^{\rm coarse})_k\,^j$ is a coefficient matrix with integer entries.

However, as described in~\sect{sect:PrimitiveAmpls}, following
ref.~\cite{TwoQuarkThreeGluon} for the case of a single fermion line,
when fundamental quarks are present these objects can be split further
into finer gauge invariant pieces. These pieces are distinguished by
the direction of the fermion line relative to the loop. Following the
fermion line into the diagram, one defines `left-turner' (L) fermions
as those which pass to the left of the loop, while `right-turner' (R)
fermions pass to the right. If the loop is in fact a closed fermion
loop, the diagram acquires an additional label $n_f$. These concepts
have already been discussed in \sect{sect:PrimitiveAmpls}. We extend
this grouping of diagrams to generic processes by labelling each
fermion line as a left- or right- turner. We thus take the primitive
amplitudes to be
\be \label{eq:setofprims} A^{D_1\dots D_{N_q}}(1,\dots,n), \ee
where $N_q$ is the number of fermion lines and $D_i \in \{L,R\}$
labels each as either a left or right turner. In the argument list we
have suppressed the identities of the $2N_q$ fermions and $n-2N_q$
gluons.

Let us suppose there are $N_{\rm prim}$ primitive amplitudes. Using our 
setup we derive an expression for each of them as a sum of diagrams,
just as we did for the partial amplitudes in \eqn{partialamp},
\be \label{primitiveamp} A_k^{{\bf D}_k} \equiv A^{{\bf D}_k}(\sigma_k) =
\sum_{j=1}^{N_d} L_k\,^j\, d_j, \qquad k=1,\ldots, N_{\rm prim} 
\ee
where the direction labels are implied. Here $L_k\,^j$ is a $N_{\rm prim} \times
N_d$ matrix describing how the j'th diagram contributes to the k'th
primitive. 

We now attempt to express the partial amplitudes of
\eqn{partialamp} as linear combinations of the primitive amplitudes of
\eqn{primitiveamp}. We write
\be \label{tosolve} A_i^{\rm partial} = \sum_{k=1}^{N_{\rm prim}} Z_i\,^k\,
A_k^{{\bf D}_k}, \ee
where $i$ runs from $1$ to $N_P$, and look for solutions for the coefficients
$Z_i\,^k$ of this set of linear equations,
\be K_i\,^j=Z_i\,^k\,\,L_k\,^j\,.\label{eqn:finalequations} \ee
The solutions of this set of equations are typically not unique as
will be discussed further in \sect{sect:relations}.

We have confirmed that the solutions match those obtained from the
known all-multiplicity expressions in ref.~\cite{Neq4Oneloop} in the
case of purely gluonic processes, and ref.~\cite{TwoQuarkThreeGluon}
for processes with one quark line. For processes with more than one
quark line we reproduce the closed form analytic expressions available
in refs.~\cite{ZqqQQ,Zqqgg} and \cite{W3jcolordec}. Beyond that we are
able to present new results as presented in more detail
in~\sect{Results}.

\subsection{Relations}
\label{sect:relations}

The linear equations (\ref{eqn:finalequations}) typically give rise to
non-trivial solutions $\{(Z^{\rm rel})^k\}$,
\be (Z^{\rm rel})^k\,\,L_k\,^j=0\,.\label{eqn:kernel} \ee
This means that the Feynman diagrams associated with a linear
combination of primitive amplitudes add up to zero.  Interpreted for
the primitive amplitudes this implies linear relations,
\be \label{relations} 0=\sum_{k=1}^{N_{\rm prim}} (Z^{\rm rel}_i)^k\, A_k^{{\bf
D}_k}\,,\quad i=1,\ldots,N^{\rm relations}\,, \ee
for the case that $N^{\rm rel}$ solutions $(Z^{\rm rel}_i)^k$ have been found.
In order to verify the relations no detailed knowledge of the Feynman
amplitudes is necessary. They arise at the diagrammatic level when taking
into account the anti-symmetry of colour-ordered three-point vertices.

It is tempting to relate the above relations to the ones that are
known for colour-ordered tree amplitudes, such as the $U(1)$
decoupling identity~\cite{MPXColor,BGColor,BGngluons} and non-abelian
generalizations~\cite{BGrelations} through the Kleiss-Kuijf
relations~\cite{KKrelations}. The role of these relations for
loop-level colour-decomposition has already been pointed out some time
ago~\cite{Neq4Oneloop,UnitarityMethod,DDMcolor} and recently reviewed in
ref.~\cite{EKMZreview}.  We will not attempt a detailed analysis into
this direction but make some general remarks about the observed
relations~(\ref{relations}).

One remark concerns the importance of having multiple fermion lines to
obtain the above relations. When considering fermion lines the
anti-symmetry of the fermion-fermion-gluon three point vertex implies
certain fermionic amplitudes may be related. The complete list of all
relations is a by-product of our way to obtain the colour
decompositions of partial amplitudes.  By inspection we observed that
the relations in \eqn{relations} originate from this anti-symmetry.
We will discuss two examples below in \sect{sect:examplerel}.  As the
multiplicity of quark amplitudes is increased by addition of gluons,
the number of relations increases due to the different ways gluons can
be added to ordered amplitudes. The new relations appear as
descendants of the ones of the purely fermionic amplitudes.

A second remark is about an application of the above relations. For
numerical evaluation of scattering amplitudes (see later
in~\sect{Results}) we exploit the relations between primitive
amplitudes (\ref{relations}) to optimise caching and thus run-times.
In the explicit expressions given, we pick methodically a subset of
primitive amplitudes using the relations. In fact we sort primitive
amplitudes according to the number of propagators of their parent
diagrams. We then apply the relations to eliminate primitive
amplitudes starting from the ones with the minimal number of parent
propagators as part of the loop.  This procedure minimizes the number
of primitive amplitudes that need to be computed. However, it does not
take into account which primitive amplitudes are most easily computed.

Similarly, one could make use mostly of primitive amplitudes with a
minimal number of parent loop propagators. For a unitarity-based
algorithm this choice would be beneficial, reducing the number of
unitarity-cuts to consider.  We leave a more thorough analysis of the
relations between primitive amplitudes to the future.  It seems likely
that an understanding of the relations between primitive amplitudes
(\ref{relations}) will be helpful towards establishing all-$n$
formulae for the colour decompositions, which are beyond the scope of
the present article.

\subsubsection{Example Relations}
\label{sect:examplerel}
\begin{figure}
  \includegraphics[scale=0.4]{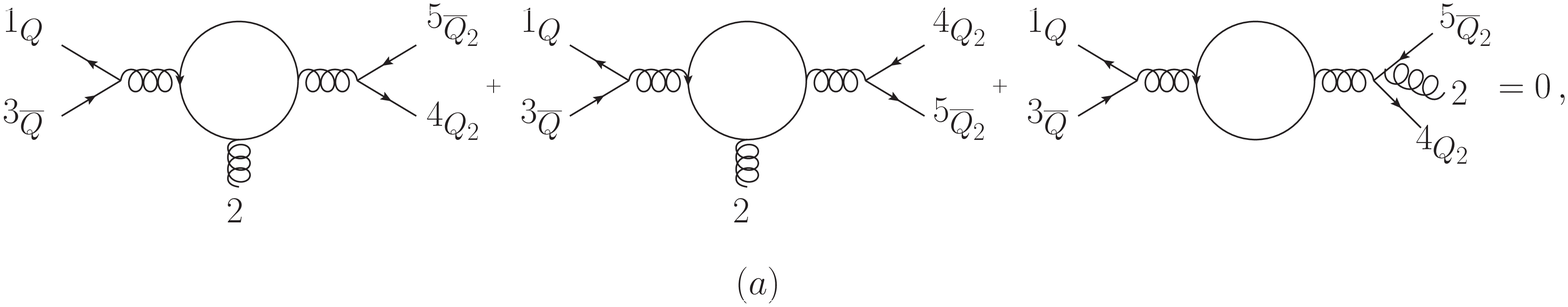}\\
  \vskip 0.4cm \includegraphics[scale=0.4]{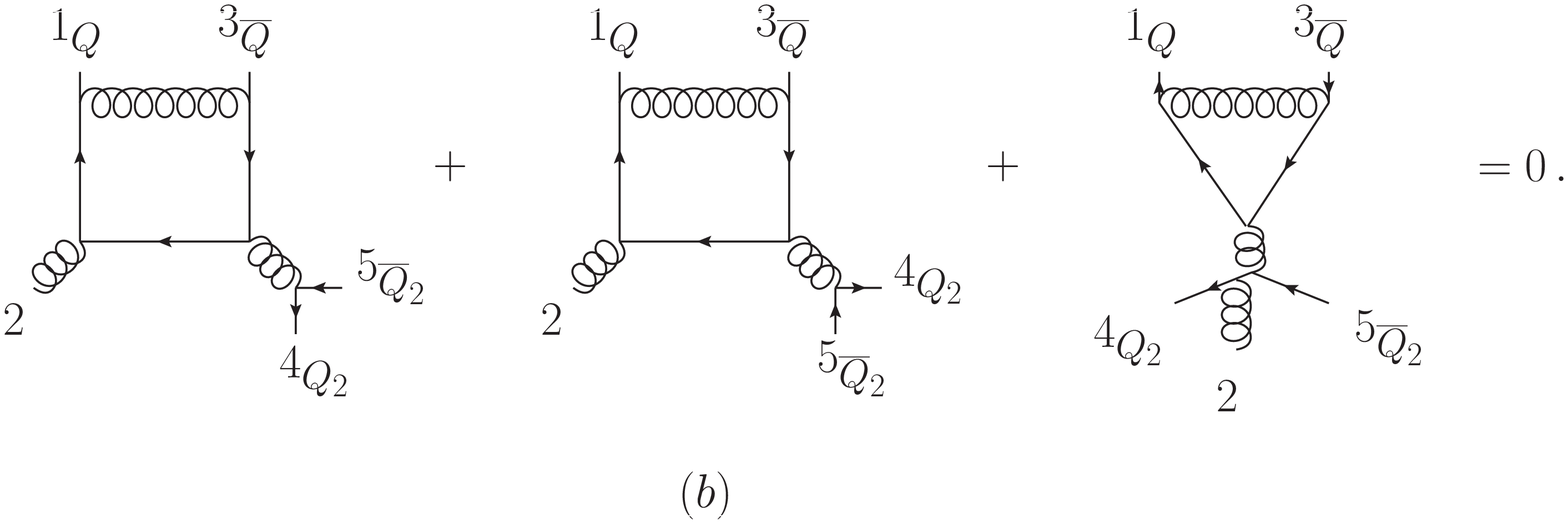}
\caption{
Parent diagram representation of relations between multi-quark primitive
amplitudes: Diagrams (a) show a vanishing sum of primitive amplitudes
associated to parent diagrams with a closed fermion loop, as in
\eqn{eq:relation_nf}.  Diagrams (b) show a vanishing sum of primitive
amplitudes with at least one gluon in the loop, the pictorial form
of~\eqn{eq:relation}.  The relations originate from exchanging the ordering of
the fermion pair $\{4_{\Q{2}},5_{\Qb{2}}\}$,which gives a relative sign.
Parent diagrams represent classes of colour-ordered Feynman diagrams.  The
contributions of all colour-ordered Feynman diagrams cancel once the asymmetry
of the fermion-fermion-gluon vertices is taken into account. }
\label{fig:relation} 
\end{figure}

For the set of primitive amplitudes we are dealing with the origin of these
relations are the symmetry properties of the quark-quark-gluon vertices. A
simple example with an internal fermion line is given by,
\be\label{eq:relation_nf}
A^{\rm n_fLL}(1_Q,5_\Qb{2},4_\Q{2},2,3_\Qb{})+
A^{\rm n_fLR}(1_Q,4_\Q{2},5_\Qb{2},2,3_\Qb{})+
A^{\rm n_fLL}(1_Q,5_\Qb{2},2,4_\Q{2},3_\Qb{})=0\,,\\
\ee
with the parent diagrams shown in~\fig{fig:relation}. Similarly, relations my
be found for diagrams with no closed internal fermion line,
\be\label{eq:relation} 
A^{\rm LL}(1_Q, 3_\Qb{}, 5_\Qb{2}, 4_\Q{2}, 2) 
+A^{\rm LR}(1_Q, 3_\Qb{}, 4_\Q{2}, 5_\Qb{2}, 2) 
+A^{\rm LL}(1_Q, 3_\Qb{}, 5_\Qb{2}, 2,
4_\Q{2})=0\,, 
\ee 
with the parent diagrams shown in~\fig{fig:relation} (b).

For both diagrammatic equations, (a) and (b) in~\fig{fig:relation},
the same mechanism is at work. Contributions from the first two
diagrams, respectively, with a direct
($4_\Q{2}$,$5_\Qb{2}$,gluon)-vertex cancel against the ones with a
direct ($5_\Qb{2}$,$4_\Q{2}$,gluon)-vertex, given that the exchange of
quarks $4_{Q_2}\leftrightarrow 5_{{\bar Q}_2}$ introduces a relative
minus-sign.  The remaining contributions include Feynman diagrams with
the gluon moved onto the fermion line. For the first parent diagram in
in~\fig{fig:relation} that would be gluon emission from $4_\Q{2}$, and
for the second parent diagram emission from $5_\Qb{2}$. These
contributions are distinct and it is the role of the Feynman diagrams
represented by the third parent diagram, to cancel off these remaining
pieces.

\subsection{A four-point example}
\label{Example}

\begin{figure}
  \includegraphics[scale=0.7]{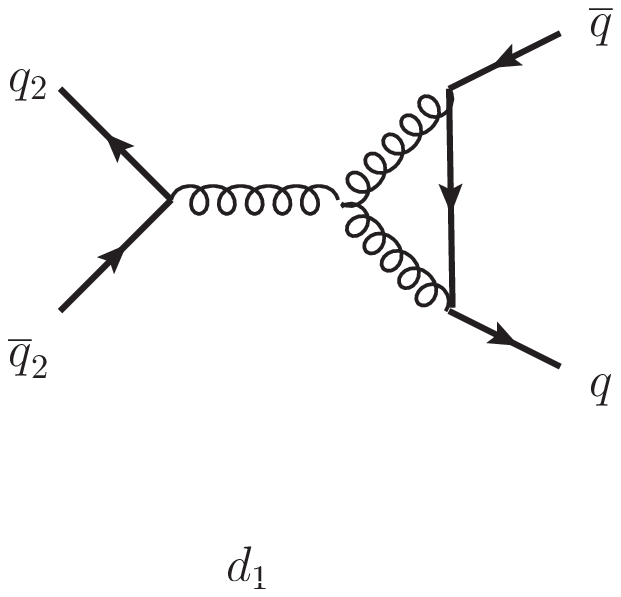}
  \includegraphics[scale=0.7]{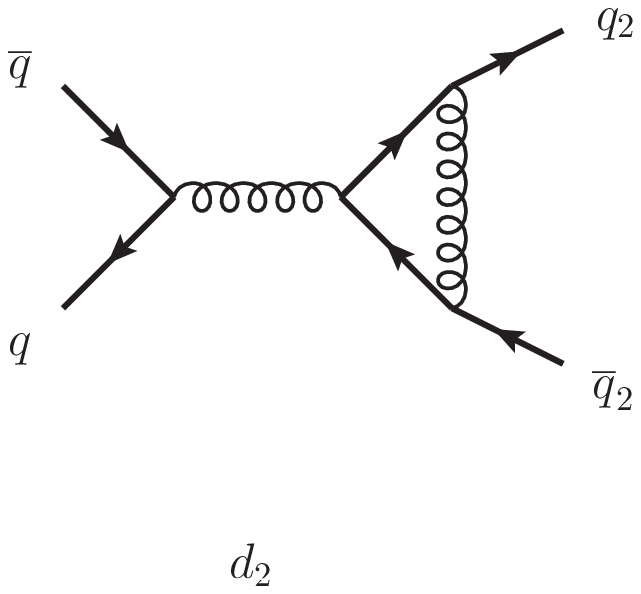}
  \includegraphics[scale=0.7]{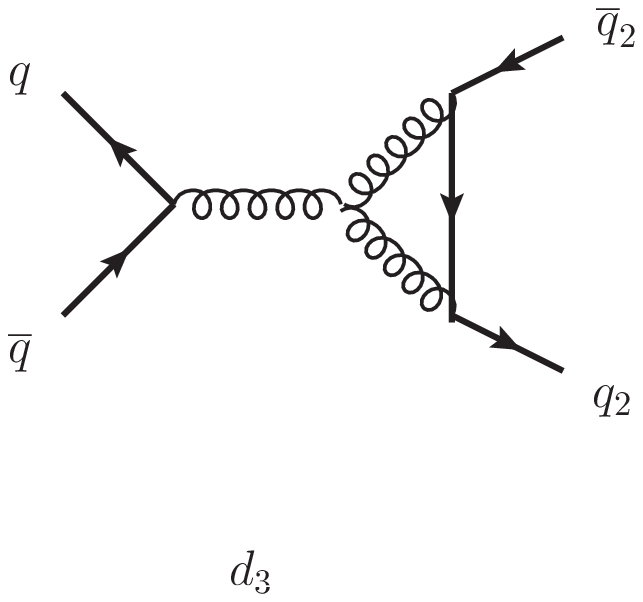}
  \includegraphics[scale=0.7]{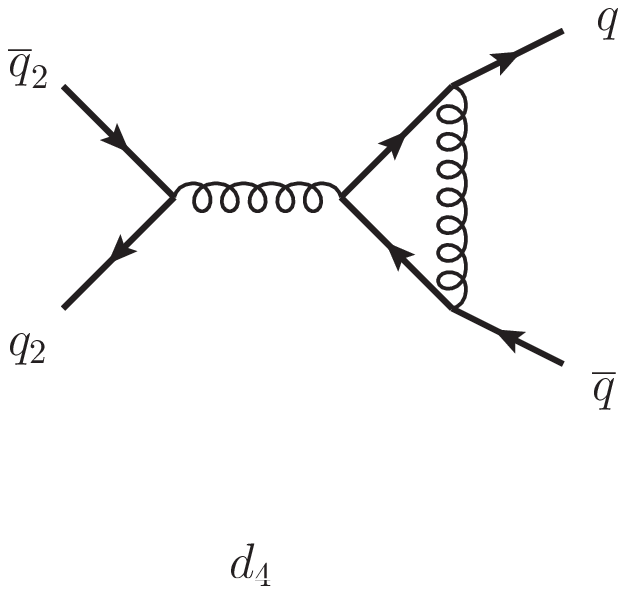}
  \includegraphics[scale=0.7]{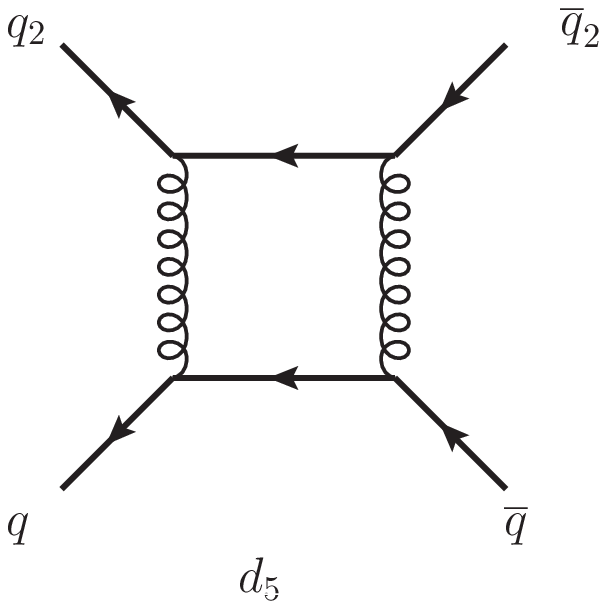}
  \includegraphics[scale=0.7]{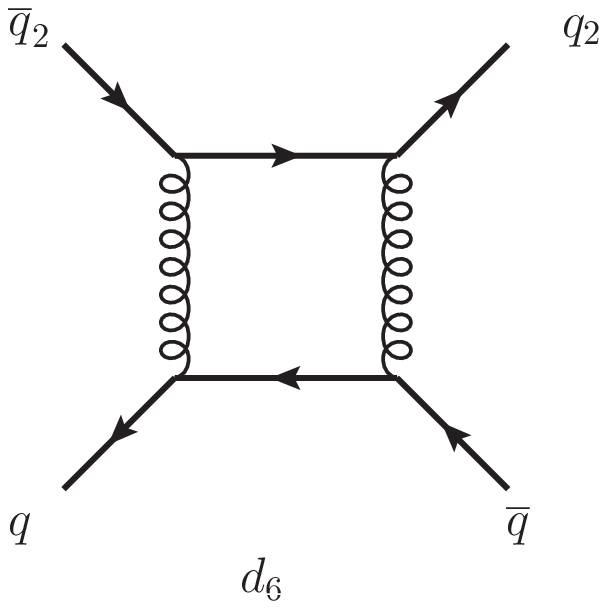}
\caption{Feynman diagrams $d_i$ contributing to the four-quark amplitude. For simplicity
we consider only box and triangle diagrams. These are sufficient to obtain the
colour decomposition of the four-quark partial amplitudes. } \label{example_FDs}
\end{figure}

To illustrate the method described in the previous section we present
here a fully worked example. We consider the amplitude for scattering
of two pairs of different flavour quarks,
\be
\mathcal{A}(1_q,2_\qb{},3_\q{2},4_\qb{2})\,.
\ee

To keep things as simple as possible we will ignore contributions with
closed fermion loops, though in our final results these pieces are of
course included.  We make one further simplification for transparency
of the example: we consider six Feynman diagrams ($N_d=6$),
illustrated in figure \ref{example_FDs}, and drop bubble
diagrams. These diagrams are sufficient for the present example.
After dressing them with colour (quarks and anti-quarks live in the
fundamental and anti-fundamental representations of $SU(N_c)$
respectively), and simplifying using \eqn{eq:fierz}, we find two
distinct colour structures ($N_P=2$),
\be 
C_1 = \delta_{i_1}^{\ib_2} \delta_{i_3}^{\ib_4}, 
\qquad 
C_2 = \delta_{i_1}^{\ib_4} \delta_{i_3}^{\ib_2}. 
\ee
The amplitude is a linear combination of $C_1$ and $C_2$,
\be 
\mathcal{A}(1_q,2_\qb{},3_\q{2},4_\qb{2}) = 
A_1^{\rm partial}(1_q,2_\qb{};3_\q{2},4_\qb{2})\, C_1 
+ 
A_2^{\rm partial}(1_q,4_\qb{2};3_\q{2},2_\qb{})\,C_2
\,. 
\ee
The expression of the partial amplitudes, 
$A_1^{\rm partial}(1_q,2_\qb{};3_\q{2},4_\qb{2})$ 
and
$A_2^{\rm partial}(1_q,4_\qb{2};3_\q{2},2_\qb{})$, 
in terms of the Feynman diagrams in figure \ref{example_FDs} is described in
(\ref{partialamp}) through the $2 \times 6$ matrix of coefficients $K_i\,^j$,
%
{\setlength\arraycolsep{2.5mm} 
\be 
\{K_i\,^j\} = \left( \begin{array}{cccccc}
-1 & \frac{1}{N_c^2} & -1 &   \frac{1}{N_c^2} & 1+ \frac{1}{N_c^2} & \frac{1}{N_c^2} 
\\[5pt] 
N_c & -\frac{1}{N_c} & N_c & -\frac{1}{N_c}  & -\frac{2}{N_c} & N_c - \frac{2}{N_c} 
\end{array} \right),  
\ee }
when contracted with the vector of Feynman diagrams $d_j$. The top row of this expression 
relates to $C_1$, and the bottom row to $C_2$.

To find the primitives, we must dress the amplitude with colour
assuming all particles are in the adjoint representation, and then
extract the primitives as the coefficients of the single trace
structures. We find
\bea \label{exampleprims} 
A^{LL}(1_Q,2_\Qb{},4_\Qb{2},3_\Q{2}) &=& -d_4, \label{eqn:prim1}\\
A^{LL}(1_\Q{},3_\Q{2},4_\Qb{2},2_\Qb{}) &=& -d_2, \\
A^{LL}(1_Q,4_\Qb{2},3_\Q{2},2_\Qb{}) &=& d_1+d_3+d_6,\\
A^{LR}(1_\Q{},2_\Qb{},3_\Q{2},4_\Qb{2}) &=& d_4, \\
A^{LR}(1_\Q{},3_\Q{2},4_\Qb{2},2_\Qb{}) &=& d_5-d_3-d_1,\\
A^{LR}(1_\Q{},4_\Qb{2},3_\Q{2},2_\Qb{}) &=& d_2.  \label{eqn:primexample}
\eea
Already at this level we observe that the primitive amplitudes are related as,
\bea
A^{LL}(1_Q,2_\Qb{},4_\Qb{2},3_\Q{2}) &=&
-A^{LR}(1_\Q{},2_\Qb{},3_\Q{2},4_\Qb{2})\,,\nonumber\\ \quad
A^{LL}(1_\Q{},3_\Q{2},4_\Qb{2},2_\Qb{}) &=&
-A^{LR}(1_\Q{},4_\Qb{2},3_\Q{2},2_\Qb{})\,.  \label{eqn:relexample}
\eea
These relations are an example of the ones discussed in~\sect{sect:relations}.
They are not direct consequences of the symmetry properties of the primitive
amplitudes (\ref{eqn:cyclsymm}), (\ref{eqn:flipsymm}) and
(\ref{eqn:flipfermarrow}). Rather they are explained by the anti-symmetry of
the colour-ordered fermion-fermion-gluon vertex. In \eqns{eqn:relexample} the
exchange $3\leftrightarrow4$ and $LL\leftrightarrow LR$ relates the primitives
up to a sign. This can also be inferred from the diagrams $d_2$ and $d_3$ in
\fig{example_FDs}.
Following our earlier notation (\ref{primitiveamp}) the equations
(\ref{eqn:prim1})-(\ref{eqn:primexample}) are expressed through a matrix of
coefficients $L_k\,^j$ and a vector of primitives $A^{D_k}_k$,
\be \label{Lmatrix} 
\{L_k\,^j\} = \left( \begin{array}{cccccc} 
0 & 0 & 0 & -1 & 0 & 0 \\ 
0 & -1 & 0 & 0 & 0 & 0 \\ 
1 & 0 & 1 & 0 & 0 & 1 \\ 
0 & 0 & 0 & 1 & 0 & 0 \\ 
-1 & 0 & -1 & 0 & 1 & 0 \\ 
0 & 1 & 0 & 0 & 0 & 0 
\end{array} \right)\,,\quad
\{A^{D_k}_k\}=\left( \begin{array}{c}
A^{LL}(1_Q,2_\Qb{},4_\Qb{2},3_\Q{2})\\
A^{LL}(1_\Q{},3_\Q{2},4_\Qb{2},2_\Qb{})\\
A^{LL}(1_Q,4_\Qb{2},3_\Q{2},2_\Qb{})\\
A^{LR}(1_\Q{},2_\Qb{},3_\Q{2},4_\Qb{2})\\
A^{LR}(1_\Q{},3_\Q{2},4_\Qb{2},2_\Qb{})\\
A^{LR}(1_\Q{},4_\Qb{2},3_\Q{2},2_\Qb{})
\end{array} \right)\,.
\ee
We note that $L$ is generally not square, though in this case it is
because we chose a reduced number of Feynman diagrams, which happens
to be equal to the number of primitive amplitudes.

The solution to equation (\ref{tosolve}) can be found by eliminating the $d_i$
from eqns.~(\ref{partialamp}) using linear combinations of the primitive
amplitudes (\ref{primitiveamp}).  We obtain the coefficient matrix $Z_i\,^k$ by
solving the linear system of equations (\ref{eqn:finalequations}) with {\tt
Mathematica}~\cite{mathematica}.
 
It turns out that there are many different solutions, reflecting the relations
(\ref{eqn:relexample}) between primitive amplitudes.  This is a new feature for
multiple quark line processes that is not observed in the purely gluonic, or
single quark line cases. In fact, the matrix $\{L_k\,^j\}$ is a $6\times6$
matrix with rank four. The associated null space (\ref{eqn:kernel}) is
two-dimensional and given by the two vectors $(Z^{\rm rel}_i)^k$ with $i=1,2$.
In matrix notation we have,
\be 
\label{example_rel} 
(Z^{\rm rel}_i)\,^k = \left( \begin{array}{cccccc} 1 & 0& 0& 1 & 0 & 0\\ 
	0& 1 & 0 & 0 & 0 & 1 
\end{array} \right). 
\ee
The null vectors imply the same relations given already in
\eqns{eqn:relexample}.

The solutions are represented as a $N_P \times
N_{\rm prim}=2 \times 6 $ matrix $Z$ as follows
\be \label{solution} Z_i\,^k = \left( \begin{array}{cccccc}
-\frac{1}{N_c^2}+\alpha &-\frac{1}{N_c^2} + \beta & \frac{1}{N_c^2} & \alpha & 1+\frac{1}{N_c^2} & \beta 
	\\ 
\frac{1}{N_c}+\gamma & \frac{1}{N_c} + \delta & N_c - \frac{2}{N_c} & \gamma & -\frac{2}{N_c} & \delta 

\end{array} \right), \ee
where $\alpha$, $\beta$, $\gamma$ and $\delta$ are arbitrary complex
numbers parameterising the solution space. The first row of $Z$ gives
the six coefficients of the primitive amplitudes, in the basis
(\ref{exampleprims}), needed to construct the coefficient of the
colour structure $C_1$. The second row does the same thing for the
structure $C_2$.

Anticipating the explicit results in~\sect{Results} we relate the solutions
(\ref{solution}) to the attached data-files.  Compared to the explicit colour
decompositions given in the attached file (partials\_4q.dat) we have the
following choices for the parameters, 
$\alpha=0$, $\beta=\frac{1}{N_c^2}$, $\gamma=0$, and $\delta=-\frac{1}{N_c}$.
In this way we avoid the use of two out of six primitive amplitudes.

\section{Results}
\label{Results}

General expressions for the decomposition of partial amplitudes in terms of
sets of primitive amplitudes are available in the literature for two classes of
process, the $n$-gluon~\cite{BKColor,Neq4Oneloop} and the two-quark
$n$-gluon~\cite{TwoQuarkThreeGluon} QCD amplitudes.
A fixed-multiplicity decomposition of four-quark amplitudes was given
in~\cite{ZqqQQ,Zqqgg} and for an additional gluon in~\cite{W3jcolordec}. In
fact these expression were first given including particles neutral under the
gauge group.  Here we generate sets of partial amplitudes algorithmically for a
given scattering process with emphasis on the cases of six- and seven-parton
QCD amplitudes including four quarks and six quarks. The generalisation of
these to amplitudes including a single colourless vector-boson is
straightforward, and will be discussed in \sect{sect:application_W4j}. In the
recent results \cite{W4jets,Z4jets} for $V$-boson plus four-jet NLO cross
sections, a leading-colour approximation was used in the virtual contribution.
In the following we will present a corresponding full-colour distribution.

\subsection{New colour decompositions for QCD amplitudes}

We present first our results for partial amplitudes in terms of primitive 
amplitudes with six and seven coloured partons. We focus entirely on processes
with distinct quark flavours. This incurs no loss of generality, because all
other cases can be derived as simple linear combinations of these.

As an example we show in \fig{textExample6q} one of the partial
amplitudes relevant for six-quark processes. In this case there are 26
contributing primitives, each with its own $N_c$-dependent
coefficient. There are three turner labels, one for each quark
line. As described in~\sect{Setup}, when there are
multiple quark lines present the expressions are not unique, as the
primitive amplitudes satisfy a set of relations.
\begin{figure} 
\includegraphics[scale=1.7,angle=-90]{diagrams/partial_6q.dat.epsi}
\caption{
A six-quark partial amplitude given as a decomposition into primitive
amplitudes. The shown partial amplitude is an extract from the file ${\rm partials\_6\,q.dat}$ in the ancillary material~\cite{partialsdata}.
} 
\label{textExample6q} \end{figure}

It is interesting to consider in comparison the leading-colour part of
the above partial amplitude, as shown
in~\fig{textExample6q_LC}. Formally, the leading colour part of the
partial amplitude is here defined as the leading terms in the limit
$N_c \gg 1$ and $n_f \gg 1$ with the ratio $N_c/n_f$ fixed.
\begin{figure}
	\includegraphics[scale=1.7,angle=-90]{diagrams/partial_6q_LC.dat.epsi}
\caption{Leading colour six-quark partial amplitude given as a decomposition
into primitive amplitudes.} 
\label{textExample6q_LC} \end{figure}
The number of contributing primitive amplitudes is strongly reduced
when subleading in $N_c$ contributions are removed. This is one of the
reasons for using this approximation when constructing NLO partonic
Monte Carlo programs --- the leading-colour amplitude can be
numerically evaluated much faster. 

\begin{figure} 
\includegraphics[scale=0.5]{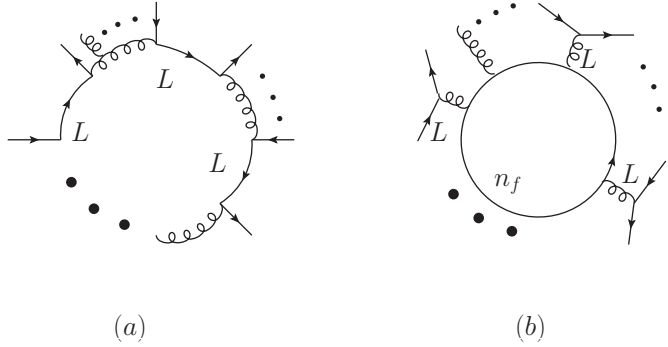}
\caption{Generic leading-colour parent diagrams for the leading partial
amplitudes: (a) a $(LL\cdots L)$-turner and (b) a $(n_fLL\cdots L)$-turner.  External
gluons may be added if directly emitted from the gluonic loop propagators in
(a) or from the fermion loop in (b).}
\label{fig:LCparents} \end{figure}

The leading-colour contribution to the decomposition of quark amplitudes is
easily obtained for arbitrary multiplicities following a few simple rules: 1)
partial amplitudes with vanishing or subleading (in $1/N_c$) tree-level
contributions are set to zero.  2) the particle orderings of the leading-colour
primitive amplitudes follow that of the born amplitude. 3) Two particular types
of left-right-turner labels are singled out leading to the two types of parent
diagrams as shown in~\fig{fig:LCparents}. In (a) we show the primitive
amplitude with multiple left-turner quark lines including gluon propagators in
the loop.  It is denoted by the label $(LLL\ldots)$. The associated parent
diagram has both gluon and fermion propagators.  The number of propagators
matches the multiplicity of the amplitude, here meant to be $n$. The second
primitive amplitude, (b), has a closed fermion loop with each fermion line
coupled to it via gluons. The label specifying the fermion routing is then
$(n_fLLL\ldots)$. The associated parent diagram has only fermion propagators.
The general structure is manifest in the two-, four- and six-quark cases
discussed here.

\subsection{Explicit results in attached text files}
\label{sect:datafiles}

The expressions for the partial amplitudes can be found in the
location~\cite{partialsdata} in the respective text files
\be {\rm partials\_\,n_g\,g\,n_q\,q.dat} \ee
provided for all independent partial amplitudes with $n_g$ gluons and $n_q$
quarks.  The format of the files has been described
in~\sect{sect:PartialNotation}.  We give the cases, 
\be \{(n_g,n_q)\}=\{(0,4),(1,4),(2,4),(3,4),(0,6),(1,6)\}.  \ee 
Together with the $(n,0)$ and $(n-2,2)$ assemblies given in
refs.~\cite{Neq4Oneloop,TwoQuarkThreeGluon}, this completes the colour
decomposition of QCD amplitudes up to and including seven colour-charged
external states.

We confirmed that two independent implementations of the algorithm
described in this paper produce identical numerical results for
partial amplitudes. The explicit expressions from each implementation
match up to application of the relations described in~\sect{sect:relations}.
Furthermore, the expressions for the partial amplitudes are checked
against the existing results in the literature where available.  In
particular, for all new six-parton and seven-parton expressions we
confirmed numerically that the virtual squared matrix elements give
the expected pole structure~\cite{CataniIRPoles} for single and double
poles in the dimensional regularization parameter $\epsilon$.  We used
\BlackHat{}~\cite{BlackHatI} and \SHERPA{}~\cite{Sherpa} for this
numerical check. A further numerical check is provided by the recent public
release of \HELAC{}~\cite{HPP,helacnlo}, which we have used to check
all our 6-parton results.

In addition to the expressions for pure QCD amplitudes, the attached files (for
the download location see~\cite{partialsdata})
\be {\rm partials\_\,n_g\,g\,n_q\,q\,2l.dat} \ee 
give their generalisation to include a colour neutral lepton pair. These
contributions are the ones relevant for the computation of \Wjjjj{}-jet
production, as we discuss in the next section.

\subsection{Application to \Wjjjj{} jets}
\label{sect:application_W4j}

The results we have presented are crucial for the application of the
colour-ordered approach to vector-boson production in association with
multiple jets at NLO.  With the colour decomposition of the pure QCD
amplitudes one can obtain rather simply their generalisations to
include colourless states. We discuss here how to obtain the colour
decomposition of $W$+jets amplitudes, and give a comparison of
leading-colour versus full-colour differential distributions of the
virtual part of \Wjjjj{}-jet production at the LHC.

\subsubsection{Inserting colour-less states.}

Amplitudes with additional colourless objects -- leptons, vector bosons, Higgs
bosons, etc. -- can be accommodated in the decomposition of partial amplitudes
into primitive amplitudes starting from the pure QCD decomposition.

For \Wjn{}-jet production we consider the decay of the $W$-boson into
a lepton pair, $\nu_\ell$ and $\ell$.  These amplitudes can be
obtained by first computing amplitudes with a virtual photon that is
emitted from a quark $q$ and decays to a charged lepton pair
$(\ell_L,\ell_R)$. In a second step, the conversion to a lepton and a
neutrino $(\nu_{\ell\,L}\,, \ell_R)$ is done by multiplicative factors
including couplings and $W$-boson propagator terms.  Details about
this conversion are explained in refs.~\cite{Zqqgg,TreesFromN4}. With
this approach we consider amplitudes with a lepton pair coupled to a
quark pair via an off-shell photon. In fact, we do not need to
consider all such amplitudes; we can drop the emission of the photon
from an internal quark loop as these kind of emissions are forbidden
for $W$-production due to flavour non-conservation.

With the basic understanding that the lepton pair is coupled to the quark with
flavour one, it is straightforward to transcribe the pure QCD partial amplitude
into the one including a virtual photon decayed into a lepton pair, as
displayed in~\fig{textExample6q2l}.  The full scattering matrix elements can be
built from these basic amplitudes~\cite{Zqqgg}.  
\begin{figure}[th]
\includegraphics[scale=1.7,angle=-90]{diagrams/partial_6q2l.dat.epsi}
\caption{ Six-quark two-lepton partial amplitude given as a decomposition into
primitive amplitudes. The leptons couple via virtual photon exchange to the quarks
with flavour one, $Q$ and $\overline{Q}$. These partial amplitudes are building
blocks of the scattering amplitudes for $W$-boson production.}
\label{textExample6q2l} \end{figure}

\subsubsection{Full colour differential distributions.}

The transverse momentum, $p_T$, of the accompanying jets is one of the
most important observables in $W$-boson production.  For \Wjjjj-jet
production at NLO the differential $p_T$ distributions of the leading
four jets have been given in~\cite{W4jets}. In that paper a
leading-colour approximation was used for the virtual
parts\footnote{The other parts - the born and real emission - were evaluated
without any colour-based approximation.}.  Here we compute for the
first time the subleading-colour corrections and compare the size of
leading-colour and full-colour virtual contributions.  We use the same
basic setup and cuts as in~\cite{W4jets}.  The virtual contributions
are computed using on-shell methods via the \BlackHat{} package, while
\SHERPA{} is used to perform the phase-space integration. The
combination of these two programs has been used extensively for
studying hadron collider phenomenology at NLO
\cite{W3jDistributions,PRLW3BH,W4jets,Z4jets,TeVZ,BlackHatI,BHWpol,BHphoton},
and is very well tested. For simplicity we will here focus on the
$p_T$ distribution of the fourth jet.

The leading-colour approximation used here differs from the one in
ref.~\cite{W4jets} by terms subleading in $1/N_c$. Both approximations
drop only finite terms (in the $1/\epsilon$ expansion) and retain the
full colour dependence of the IR-divergent pieces. Our explicit setup
for the finite parts is given in the following.  Initially, our
virtual contributions are computed in the four-dimensional helicity
(FDH) scheme~\cite{FDHolder,FDH}, which is most convenient for the use
of helicity amplitudes.  Only in the end are the virtual amplitudes
converted to the more standard 't Hooft-Veltman scheme~\cite{HVscheme}
used as input for \SHERPA{}.  This conversion is done through a simple
shift proportional to the born matrix element (see e.g.  section six
in~\cite{Zqqgg}).  Our leading-colour approximation contains the full
scheme shift contributions, including terms subleading in $1/N_c$.  In
both approaches, the partial amplitudes include the leading terms in
the formal limit $N_c \gg 1$ and $n_f \gg 1$ with $N_c/n_f$
fixed. Here, all interference terms of the non-zero partial amplitudes
are kept.  To point out one difference to ref.~\cite{W4jets}, there,
in a more conventional approach, only the leading-colour interference
terms were kept.

\begin{figure} \includegraphics[scale=0.9]{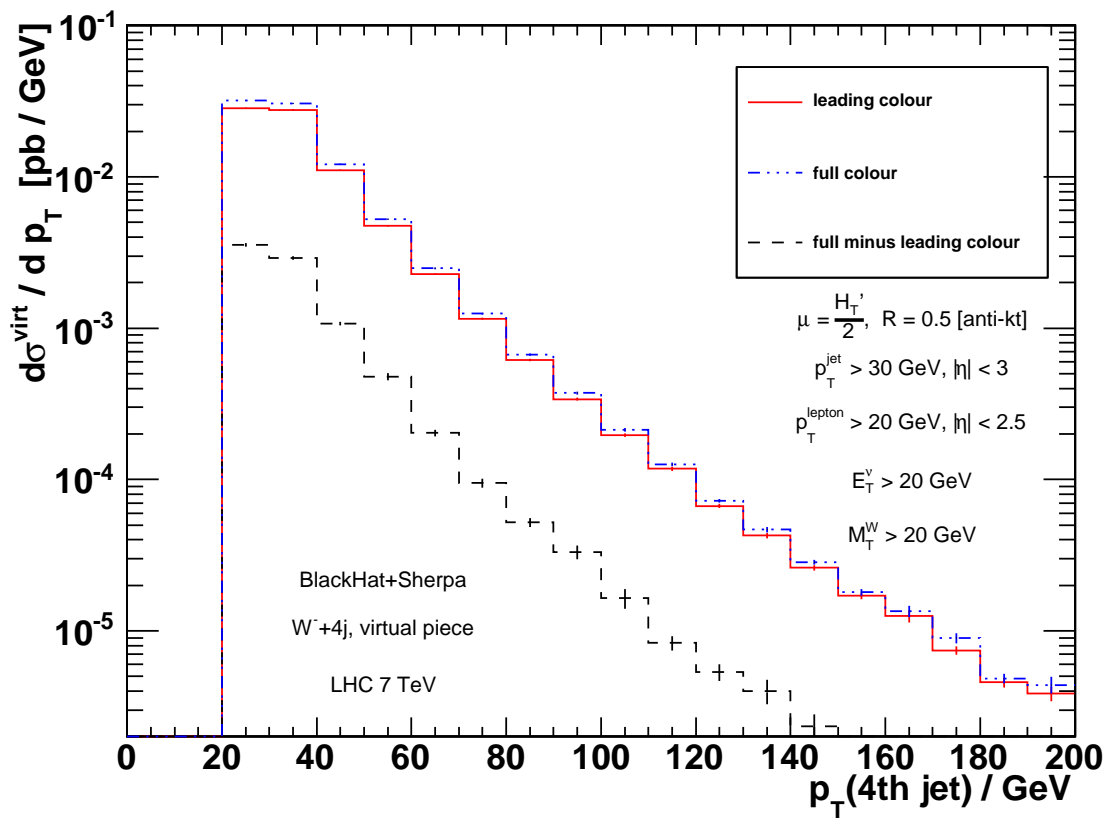}
\caption{ 
A comparison of the full and leading-colour virtual
contributions to the $p_T$ distribution of the fourth jet in
$W^-$+4-jet production at the $7$~TeV LHC. Their difference
(full-minus-leading-colour) is also shown. We show not a physical cross
section, but rather a particular piece of it. The subleading-colour
contribution is suppressed by a factor of $1/N_c^2$, approximately uniformly.
The vertical lines in the center of each bin, indicate integration errors.}
\label{Wm4j_virt_pt} \end{figure}

The comparison of the full-colour and leading-colour virtual contributions to
the $p_T$ distribution of the fourth jet in $W^-$+4-jet production at the LHC
is shown in~\fig{Wm4j_virt_pt}.  Also displayed are the subleading-colour
contributions by themselves, labelled as ``full-minus-leading-colour''.  In
order to obtain the full parton level differential cross-section one must add
the real and born contributions in the usual way.  The complete $p_T$
distribution, including born, real and leading-colour virtual contributions,
has been given already in~\cite{W4jets}.

We confirm that the subleading-colour contribution is suppressed uniformly over
the $p_T$ range from 20 GeV to 200 GeV.  The suppression appears consistent
with the expected factor of $1/N_c^2$ with $N_c=3$.  With the leading-colour
virtual part accounting for about $20\%$ of the leading-colour total cross
section, as already observed in refs.~\cite{W3jDistributions,W4jets}, the
subleading-colour virtual part amounts to less than a $3\%$ correction. 
Although numerically similar, the present leading colour
approximation and the one from ref.~\cite{W4jets} are not identical. 
We have checked that our conclusions hold for both definitions, namely that
sub-leading colour contributions are at the level of a few percent.

A possible exception to the uniform suppression are zeros of the leading-colour
virtual cross section. Such zeros are not excluded on general grounds, as the
virtual piece alone is not physical. A priori, there is no reason to assume that
the vanishing of the leading-colour contribution forces also the vanishing of
the subleading-colour contribution.  Close to a zero of the leading-colour
contribution we naturally expect a relative enhancement of the
subleading-colour piece.  In the range of the distribution presented
in~\fig{Wm4j_virt_pt} we do not observe such behaviour and observe a uniform
suppression of the subleading-colour contribution. We leave a more thorough
consideration of this issue, as well as a more detailed phenomenological
analysis, to future work.

\section{Conclusion}
\label{Conclusion}

Recent years have seen impressive progress in NLO parton level
computations relevant for LHC physics. In particular, the unitarity and
on-shell approaches have opened new possibilities for the computation
of one-loop scattering amplitudes.

An important ingredient in these advances is a clear theoretical
understanding of the amplitudes, which enables an efficient
structuring of the calculation. This progress has made possible high
multiplicity parton-level predictions at NLO for generic physics
processes and models.  The colour organisation discussed here is one
important aspect towards automation in the colour-ordered approach.

We have presented explicit expressions for the colour decomposition of
pure-QCD amplitudes with up to seven partons. We also gave their
extension to include a leptonically decaying $W$-boson. Whereas
leading-colour expressions are simple to write down, terms subleading
in $1/N_c$ are not easily obtained. We have described a general
algorithm for obtaining such subleading terms, applicable to cases
with any number of fermion pairs.

Although we give explicit expressions, typically these are not
unique. There exist relations between primitive amplitudes which
originate in the anti-symmetry of the fermion-fermion-gluon
colour-ordered three-point vertex.  We expect that these
``fermion-flip'' relations will be important in obtaining all-$n$
formulae for multi-quark amplitudes, as exist already for the
pure-gluon~\cite{Neq4Oneloop} and two-quark~\cite{TwoQuarkThreeGluon}
cases (see also~\cite{DDMcolor}).

Contributions to the virtual corrections at subleading order in $1/N_c$ have
been shown to be uniformly small over phase space for \Wjjjx{}-jet production
in~\cite{W3jDistributions}.  However, no general theorems are available to
assess the size of subleading-colour terms in all regions of phase space
\emph{a priori}.  We have applied our new results to understand the effect of
subleading-colour virtual contributions to distributions of \Wjjjj{}-jet
production.  We showed that for the fourth-jet $p_T$ distribution the effect is
small (a few percent) and observe a similar behaviour for other distributions.
A detailed study of the physics, including subleading-colour contributions, to
pure jets and vector-boson plus jets could be carried out in the future using
the results of this article.

In summary, we have presented a general method towards a colour-ordered
approach for loop computations with many coloured final states.  We expect these results will aid in physics studies of relevance to the LHC in the
near future.

\section*{Acknowledgments}

\vskip -.3 cm 

We would like to thank Zvi Bern, Giovanni Diana, Lance Dixon, Fernando
Febres Cordero, Stefan H\"oche, David Kosower and Daniel Ma\^itre for
many useful discussions and comments on the manuscript. HI would like
to thank the School of Physics and Astronomy at Tel-Aviv University
for their hospitality during the completion of this work.

This research was supported by the US Department of Energy under
contracts DE--FG03--91ER40662. HI's work is supported by a grant from
the US LHC Theory Initiative through NSF contract PHY-0705682.

\end{document}